\documentclass[final, 11pt]{article}

\usepackage{amssymb, amsmath, amsthm, mathtools}
\usepackage[usenames]{color} 
\usepackage{booktabs} 
\usepackage{multirow} 
\usepackage{graphics}
\usepackage{pdflscape} 
\usepackage{longtable} 
\usepackage{rotating} 
\usepackage{subcaption} 
\usepackage{enumerate} 
\usepackage{natbib}
\usepackage{tikz}
\usepackage{todonotes}
\usetikzlibrary{decorations.pathmorphing,calc,shapes,shapes.geometric,patterns,fit}

\usepackage{floatrow}
\usepackage[top=1.25in, bottom=1.25in, left=1in, right=1in]{geometry} 

\usepackage{setspace} 
\usepackage{hyperref} 

\newcommand{\R}{\mathbb{R}} 

\newcommand{\ind}{\mathbf{1}} 
\newcommand{\norm}[1]{\left|\left|#1\right|\right|} 
\newcommand{\smallnorm}[1]{||#1||} 
\newcommand{\abs}[1]{\left|#1\right|} 

\DeclareMathOperator*{\argmin}{arg\,min}

\theoremstyle{plain}
\newtheorem{thm}{Theorem}[section]

\newtheorem{cor}[thm]{Corollary}

\theoremstyle{definition}

\theoremstyle{remark}
\newtheorem{rem}[thm]{Remark}

\theoremstyle{plain}
\newtheorem*{empiricalimp*}{Empirical Hypothesis}

\newtheoremstyle{named}{}{}{\itshape}{}{\bfseries}{.}{.5em}{\thmnote{#3 }#1}
\theoremstyle{named}

\theoremstyle{plain}
\newtheorem*{thm*}{Theorem}
\newtheorem*{exercise*}{Exercise} 
\newtheorem*{example*}{Example} 
\newtheorem*{discussion*}{Discussion} 
\newtheorem*{claim*}{Claim}
\newtheorem*{assumption*}{Assumption} 
\newtheorem*{lem*}{Lemma}
\newtheorem*{prop*}{Proposition}
\newtheorem*{cor*}{Corollary}
\newtheorem*{defn*}{Definition}

\theoremstyle{remark}
\newtheorem*{rem*}{Remark}
\newtheorem*{note*}{Note}
\newtheorem{case*}{Case}

\usepackage{sectsty}

\usepackage{appendix} 

\usepackage[USenglish]{babel}

\DeclareCaptionFormat{cont}{#1 (continued)#2#3\par}

\title{Statistical Arbitrage Risk Premium by Machine Learning}
\author{Raymond C.\ W.\ Leung and Yu-Man Tam
  \footnote{%
	Leung is with the Cheung Kong Graduate School of Business, and Tam is with the Office of the Comptroller of the Currency, U.S.\ Department of Treasury. This paper is a substantially revised version of a paper previously circulated under the title ``Asset Insurance Premium in the Cross-Section of Asset Synchronicity''. Please send all correspondences to Leung at \href{mailto:raymond.chi.wai.leung@gmail.com}{\nolinkurl{raymond.chi.wai.leung@gmail.com}}. Part of this research is done while Leung was visiting UC Berkeley. We thank Robert M.\ Anderson, Lisa Goldberg, Alex Michaelides, Neal Stoughton, Baolian Wang, Hongjun Yan, and Wenhao Yang for helpful discussions. We also thank the seminar participants of the Five-Star Forum, Internal Machine Learning Workshop at the OCC, SWUFE-CDAR 2018 Symposium and the 2019 FMA Annual Meeting in New Orleans for helpful comments and suggestions. We thank an anonymous referee who made several suggestions that significantly improved the paper. Views and opinions expressed are those of the authors and do not necessarily represent official positions or policy of the OCC. All errors are ours.   %
	}
}
\date{March 2021}

\newcommand{\MainSpec}{20210305_1500_normedcoef_elastic_net}  

\newcommand{\MainSpecFIG}{\MainSpec/figures/}

\newcommand{\MainSpecPortSortFIG}{\MainSpec/figures/elastic_net_r2}

\newcommand{\myinput}[1]{%
	\let\center\empty
	\let\endcenter\relax
	\centering
	\resizebox{\linewidth}{!}{
    \input{#1}
}
}

\newcommand{\E}{\mathbb{E}} 
\newcommand{\Var}{\ensuremath{\mathrm{Var}}} 

\newcommand{\Rsquared}{\ensuremath{\mathsf{R^2}}} 

\def\signed #1{{\leavevmode\unskip\nobreak\hfil\penalty50\hskip2em
  \hbox{}\nobreak\hfil(#1)%
  \parfillskip=0pt \finalhyphendemerits=0 \endgraf}}
\newsavebox\mybox

\usepackage{pifont}

\begin{document}

\clearpage\maketitle
\setcounter{page}{1}
\thispagestyle{empty} 

\begin{abstract} 
  \singlespacing
  How to hedge factor risks without knowing the identities of the factors? We first prove a general theoretical result: even if the exact set of factors cannot be identified, any risky asset can use some portfolio of similar peer assets to hedge against its own factor exposures. A long position of a risky asset and a short position of a ``replicate portfolio'' of its peers represent that asset’s factor residual risk. We coin the expected return of an asset's factor residual risk as its \emph{Statistical Arbitrage Risk Premium} (SARP). The challenge in empirically estimating SARP is finding the peers for each asset and constructing the replicate portfolios. We use the elastic-net, a machine learning method, to project each stock's past returns onto that of every other stock. The resulting high-dimensional but sparse projection vector serves as investment weights in constructing the stocks' replicate portfolios. We say a stock has high (low) \emph{Statistical Arbitrage Risk} (SAR) if it has low (high) R-squared with its peers. The key finding is that ``unique'' stocks have both a higher SARP and higher excess returns than ``ubiquitous'' stocks: in the cross-section, high SAR stocks have a monthly SARP (monthly excess returns) that is 1.101\% (0.710\%) greater than low SAR stocks. The average SAR across all stocks is countercyclical. Our results are robust to controlling for various known priced factors and characteristics.
\end{abstract} 


\newpage

\emph{Given any stock, how can one hedge against its factor risks?} This question is simple to answer with a linear factor model structure. For instance, under the Markowitz mean-variance portfolio theory and its resulting equilibrium capital asset pricing model (CAPM) (\cite{Sharpe1964}, \cite{Lintner1965}), any given stock's returns can be explained by some linear combination of a risk free asset return and its beta loading on the market portfolio return. Hence, the factor risk of any stock can be hedged out by shorting its beta loading multiplied by the market factor. However, with the large number of tradable and non-tradable factors that have been documented in the literature (\cite*{Harvey2016}), our question becomes difficult to approach because its answer then significantly depends on which factors the researcher decides to include in his empirical study, and is also affected by the empirical uncertainty with estimating the factor loadings. 

The first main contribution of this paper is to \emph{theoretically} argue an effective way to hedge against potentially unidentifiable factor risks of a stock is to answer this dual question: \emph{Given any stock, what portfolio of \textit{all other stocks} is most similar to it?} Suppose all stocks are exposed to the same set of linear factors but with heterogeneous factor loadings. If one can identify a group of peers that is the most ``similar'' to a given stock $i$, then this portfolio is also exposed to similar factor loadings of this stock $i$. We view this portfolio of peer stocks as the \emph{replicate} of stock $i$. A long position on stock $i$ and a short position on its replicate will expose the holder to any remaining factor risks of stock $i$ that cannot be completely hedged out by its peer stocks. We show this long-short position exactly equates to the residual factor risks of stock $i$. Provided an econometrician has a method to find these peer stocks, this long-short position does not require the econometrician to know the true underlying factor structure of the economy. We call the expected returns of this long-short position the \emph{Statistical Arbitrage Risk Premium} (SARP) of stock $i$, and SARP is the key object of study in this paper.

How do we \emph{empirically} study SARP? Is there a cross-sectional difference in SARP? The second main contribution of the paper is answering these questions. For each month end, we use the \emph{elastic-net} estimator, a machine learning method, to project each stock $i$'s past twelve months' daily returns onto the returns of \emph{every other} stock in the market. The resulting elastic-net projection vector is high-dimensional but very sparse. After a suitable normalization, the projection vector is then used as investment weights into all stocks other than $i$. The resulting portfolio is hence a machine learning constructed \emph{replicate} of stock $i$. As theoretically motivated above, the time-series average return from a long position of stock $i$ and a short position of its replicate is the SARP of stock $i$. Moreover, we call the elastic-net projection $\Rsquared$ of each stock $i$ as the \emph{Statistical Arbitrage Risk} (SAR) of stock $i$; we say a stock has high SAR if it has a low elastic-net $\Rsquared$, while a stock has low SAR if it has high $\Rsquared$. The core empirical message of this paper can be succinctly summarized as: 
\begin{equation*}
  \text{\emph{SARP is increasing in SAR}}. 
\end{equation*}
That is in the cross-section, ``unique'' stocks (i.e.\ so having low $\Rsquared$, and hence high SAR) have a higher SARP than ``ubiquitous'' stocks (i.e.\ high $\Rsquared$, so low SAR). Over the sample period of January 31, 1976 to December 31, 2020, high SAR stocks have a monthly SARP of 1.368\% and low SAR stocks have a monthly SARP of 0.267\%, and the difference 1.101\% is highly statistically significant. And even without studying SARP, we have the important corollary that high SAR stocks have a monthly return of 1.481\% and low SAR stocks have a monthly return of 0.771\%, and the difference 0.710\% is also highly statistically significant. 

Our paper belongs to a growing literature of applying machine learning methods to study empirical asset pricing questions. Broadly speaking, many recent papers in this literature use machine learning methods for factor selection and/or forecasting. We do neither in this paper. There are only two purposes of using a machine learning method in this paper: to identify the SAR of each stock, and to construct the replicate portfolio of each stock. The estimation and inference of SARP for each stock use conventional empirical asset pricing procedures. 

Recent papers have applied variants of the \textit{least absolute shrinkage and selection operator (LASSO)} estimator of \cite{Tibshirani1996}.  \cite*{2017_WP_Feng} take advantage of the sparsity property of LASSO and develop a multi-step approach to evaluate the price of risk of a given new factor above and beyond an existing set of factors. \cite*{2018_WP_Freyberger} uses adaptive group LASSO to select characteristics that provides marginal information for the cross section of expected stock returns. The literature has documented a large set of factors or characteristics (\cite*{Harvey2016}), and it is hoped that machine learning methods can substantially shrink down the number of factors that can explain the cross-section of returns. \cite*{2018_JF_Chinco} use the LASSO to predict one-minute-ahead return using lagged high frequency returns of other stocks as regressors. \cite*{Gu2018} apply an extensive battery of machine learning methods and discover that such methods improve predictability accuracy over traditional methods. \cite{shu2020high} is a recent paper that uses an adaptive elastic-net estimator to construct a sparse portfolio that can track a large index portfolio. While both of our papers use the elastic-net estimator, our paper emphasizes the use of this estimator to discover a new asset pricing anomaly, while \cite{shu2020high} emphasizes the use of this estimator to mimic and dimension-reduce a large portfolio. 

Our paper is also related to the literature of pairs trading, substitutability of risky assets and statistical arbitrage. \cite*{Gatev2006} finds pairs of similar stocks using the minimal distance between normalized historical prices, and argue that the resulting pairs trading strategy generates abnormal returns and that the source of this profit is the mispricing of close substitutes. \cite{Krauss2017} is a recent survey of the pairs trading literature. \cite{wurgler2002does} similarly also argue that stocks without close substitutes are likely to have large mispricings; the authors identify a few predefined number of similar stocks using a predefined sorting method. In contrast by using a machine learning method in this paper, the selection of a stock's risky peers is completely data driven. Indeed, the closest substitute of a given stock could potentially be hundreds of all other stocks. \cite{huck2019large} and \cite{avellaneda2010statistical} are two recent studies on statistical arbitrage.

Section~\ref{sec:TheoreticalMotivation} lays out the theoretical framework of the paper. Section~\ref{sec:Methodology} explains our estimation methodology. The main empirical results of the paper are in Section~\ref{sec:MainResults}. Section~\ref{sec:RobustnessChecks} show additional empirical robustness checks. We conclude in Section~\ref{sec:Conclusion}. We defer the details of the elastic-net procedure to Section~\ref{sec:Appendix}. All proofs to Section~\ref{sec:TheoreticalMotivation} are in the Online Supplementary Materials \cite{leung2021supplement}.

\section{Theoretical motivation}%
\label{sec:TheoreticalMotivation} 
We first prove a general theoretical asset pricing result that will guide our empirical research design. 
\begin{thm}%
	\label{thm:ThyReturnDecomp} 
	Suppose there are $N + 1$ risky assets and a single risk-free asset. Assume all of these risky assets are governed by a linear factor structure with $K$ number of factors with risky returns $\boldsymbol{F}$,
	\begin{equation}
		R_i = \alpha_i + \boldsymbol{\beta}_i^\top \boldsymbol{F} + \varepsilon_i
		\label{eq:ThyReturnDecompA} 
	\end{equation}
	and where the idiosyncratic risk $\varepsilon_i$ of the $i$th risky asset is assumed to have zero mean and is independent of $\boldsymbol{F}$ for all $i = 1, \ldots, N + 1$. Suppose there are strictly more risky assets than factors, so $N > K$.  

	Then the excess returns $R_i$ of any individual risky asset $i$ can be expressed as a linear combination of other risky asset returns as. 
	\begin{equation} 
		R_i = \mathbf{b}_i^\top \boldsymbol{R}_{-i} + \mathbf{a}_i^\top \boldsymbol{\Phi}_i - \mathbf{b}_i^\top \boldsymbol{\varepsilon}_{-i} + \varepsilon_i. %
		\label{eq:ThyReturnDecompB} 
	\end{equation}%
	That is, the excess returns $R_i$ of any individual asset $i$ can be expressed as a combination of: (i) the $N \times 1$ vector of excess returns of \emph{all other} of risky assets $\boldsymbol{R}_{-i}$; (ii) the factor loadings on some $K$ risky asset returns $\boldsymbol{\Phi}_i$, and we will call these $K$ assets the \emph{factor residuals of asset $i$}; (iii) the $N \times 1$ vector of idiosyncratic risks $\boldsymbol{\varepsilon}_{-i}$ of all other $N$ risky assets; and (iv) the idiosyncratic risk $\varepsilon_i$ of asset $i$ itself. 

	The $N \times 1$ vector $\mathbf{b}_i$ is dependent on the intercepts $\{ \alpha_j \}_{j=1, j \neq i}^{N+1}$ and the entire factor loadings $\{ \boldsymbol{\beta}_j \}_{j=1, j \neq i}^{N+1}$ of the economy and whose analytical expression is in the Internet Appendix, and where $\mathbf{a}_i^\top := [\alpha_i, \boldsymbol{\beta}_i^\top ]$. 
\end{thm}%
This result tells us given \emph{any} factor structure in the financial markets like \eqref{eq:ThyReturnDecompA}, of which its theoretical existence can always be justified via \cite{Ross1976}, the returns of a single risky asset $R_i$ can be expressed as a linear combination $\mathbf{b}_{i}$ of returns of \emph{all other} risky assets $\boldsymbol{R}_{-i}$ like \eqref{eq:ThyReturnDecompB}. The key intuition of this result is a hedging and replication argument. Suppose the researcher does \emph{not} know the exact identifies of the factors $\boldsymbol{F} = [F_1, \ldots, F_K]^\top$ in the economy. But as long as all risky assets have an exposure to these factors, then any particular risky asset $i$ can use some combination $\mathbf{b}_i$ of other risky stocks to hedge against risky asset $i$'s factor risks.
\footnote{ 
Some readers may think that the result in Theorem~\ref{thm:ThyReturnDecomp} and the subsequent Corollary~\ref{cor:AssetInsurance} can be trivially derived by ``inverting the factor loadings'' in \eqref{eq:ThyReturnDecompA} to derive \eqref{eq:ThyReturnDecompB}. However, we should note that as there are (significantly) more assets $N$ than the number of factors $K$ in this economy. Thus the economy-wide factor loading matrix is necessarily of dimensions $N \times K$, meaning the matrix is not square and thus not invertible. We take extra care in the proofs to make sure that a proper sense of ``factor inversion'' is possible in this general setup. 
}
For the empirical component of our paper, this means projecting one stock's returns onto the returns of all other stocks is not a naive statistical exercise but actually has concrete microeconomic foundations. 

The next result will tell us the economic content of Theorem~\ref{thm:ThyReturnDecomp}. 
\begin{cor}
	\label{cor:AssetInsurance} 
	Suppose the conditions of Theorem~\ref{thm:ThyReturnDecomp} hold. For each fixed risky asset $i$, we can find $K + 1$ risky assets whose returns are given by 
	\begin{subequations} 
			\begin{align}
				\Phi_{i,0} &= 
				\begin{cases} 
					1, & \text{if $\alpha_j = 0$ for all $j \neq i$} \\ 
					1 - \frac{ \boldsymbol{\alpha}_{-i}^\top }{ \norm{ \boldsymbol{\alpha}_{-i} }_2^2} ( \boldsymbol{R}_{-i} - \boldsymbol{\varepsilon}_{-i} ), & \text{if otherwise} 
				\end{cases} 
				\label{eq:InsuranceAssetRetIntercept} \\ 
				\Phi_{i,k} &= F_k - \mathbf{c}_{i,k}^\top (\boldsymbol{R}_{-i} - \boldsymbol{\varepsilon}_{-i}), \quad k = 1, \ldots, K
				\label{eq:InsuranceAssetRet} 
			\end{align}
	\end{subequations} 
	where $\mathbf{c}_{i, k}$ is some $N \times 1$ deterministic vector such that only depends on the factor loadings $\{ \boldsymbol{\beta}_j \}_{j=1, j \neq i}^{N+1}$ and intercept $\{ \alpha_j \}_{j =1, j \neq i}^{N+1}$ structure of the economy (its analytical expression is in the Internet Appendix). Here we denote $\boldsymbol{\alpha}_{-i}^\top := [ \alpha_1, \ldots, \alpha_{i-1}, \alpha_{i+1}, \ldots, \alpha_{N+1}]$ and $\norm{\cdot}_2$ is the Euclidean norm on $\R^N$. We will denote $\boldsymbol{\Phi}_i^\top := [ \Phi_{i,0} , \Phi_{i,1}, \ldots, \Phi_{i,K} ]$.
\end{cor} 

Corollary~\ref{cor:AssetInsurance} is what motivates us to refer to those $K$ risky assets with returns $\boldsymbol{\Phi}_i$ as \emph{factor residuals for asset $i$}. From \eqref{eq:ThyReturnDecompB}, we see the following decomposition, 
\begin{equation}
	\mathbf{a}_i^\top \boldsymbol{\Phi}_i = 
	\alpha_i \Phi_{i,0} + \boldsymbol{\beta}_i^\top \boldsymbol{\Phi}_{i,1:K}, 
	\label{eq:AssetInsPremDecomp} 
\end{equation}
where $\boldsymbol{\Phi}_{i,1:K}^\top := [ \Phi_{i,1}, \ldots, \Phi_{i,K}]$. 

The first part $\Phi_{i,0}$ adjusts for any potential mispricings in the economy. As an important special case when there is no pricing at all in the economy
\footnote{ 
It is well known that if all risky assets are mean-variance efficient then necessarily there is no mispricing in the economy. See an early discussion of this classical empirical asset pricing test in \cite*{Black1972}. 
} 
meaning $\alpha_i = 0$ and $\alpha_j = 0$ for all assets $j \neq i$, then $\Phi_{i,0} = 1$ and $\alpha_i \Phi_{i,0} = 0$. If on the other hand, when the intercepts $\alpha_i, \alpha_j$'s are generically non-zero, then the factor residuals of asset $i$ will explicitly hedge out any level of mispricing in the economy through $\Phi_{i,0}$, and asset $i$'s net mispricing exposure is $\alpha_i \Phi_{i,0} = \alpha_i - \alpha_i \boldsymbol{\alpha}_{-i}^\top (\boldsymbol{R}_{-i} - \boldsymbol{\varepsilon}_{-i} ) / \norm{ \boldsymbol{\alpha}_{-i} }_2^2 $. Secondly, the next $K$ parts $\Phi_{i,k}$ adjust for factor exposures. From the perspective of an agent who owns asset $i$, he is exposed to each of the $k = 1, \ldots, K$ factor returns through the factor loadings $\boldsymbol{\beta}_i^\top = [ \beta_{i,1}, \ldots, \beta_{i,K} ]$ of \eqref{eq:ThyReturnDecompA}. However, all the other $N$ risky assets will also have some exposure to the $k$-th factor. Suppose the agent constructs an ``artificial asset'' that loads into the return of the $k$-th factor, while shorting some combination $\mathbf{c}_{i,k}$ of the factor contributions to all the other $N$ risky assets $\boldsymbol{R}_{-i} - \boldsymbol{\varepsilon}_{-i}$. The resulting artificial asset has the returns $\Phi_{i,k}$. Hence $\Phi_{i,k}$ is precisely the residual exposure of asset $i$ to the $k$-th factor, after using the returns of all other assets to ``hedge'' as much as possible this factor risk. The ``hedging'' nature of these artificial assets motivates us to call them as ``factor residuals'' for asset $i$. The holder of asset $i$ is exposed to $K$ number of factor risks, and so he will have to construct $K$ number of these factor residuals. And since asset $i$ has factor exposure to $K$ number of factors, the agent will weigh $\beta_{i,k}$ loadings into the $k$-th factor residual return $\Phi_{i,k}$, as in the $\boldsymbol{\beta}_i^\top \boldsymbol{\Phi}_{i,1:K}$ term of \eqref{eq:AssetInsPremDecomp}. 

Let's rearrange \eqref{eq:ThyReturnDecompB} and take its expectation, 
\begin{equation}
	\E[R_i] - \mathbf{b}_i^\top \E[\boldsymbol{R}_{-i}] 
	= \mathbf{a}_i^\top \E[\boldsymbol{\Phi}_i] - \mathbf{b}_i^\top \E[\boldsymbol{\varepsilon}_{-i}] + \E[\varepsilon_i] 
	= \mathbf{a}_i^\top \E[\boldsymbol{\Phi}_i], 
	\label{eq:DiffedExpectations} 
\end{equation}
where $\E[\boldsymbol{\varepsilon}_{-i}] = \boldsymbol{0}_N$ and $\E[\varepsilon_i] = 0$ because idiosyncratic risks are not priced.  The overall term $\mathbf{a}_i^\top \E[\boldsymbol{\Phi}_i]$ in \eqref{eq:DiffedExpectations} is exactly the expected portfolio return into these $K$ factor residuals. This is why we will call $\mathbf{a}_i^\top \E[\boldsymbol{\Phi}_i]$ as the \emph{Statistical Arbitrage Risk Premium (SARP) for asset $i$}. The key objective of this paper is to empirically study SARP. 

The next result shows that, under weak economic and technical conditions, SARP is non-zero. 
\begin{cor}
	\label{cor:NonZeroAssetInsPrem} 
	The SARP of any non-redundant asset $i$ is almost surely non-zero when some of the $K$ factors are correlated. 
\end{cor} 

The following result relates the regression $\mathsf{R}$-squared to Theorem~\ref{thm:ThyReturnDecomp}. 
\begin{cor}%
	\label{cor:AssetRSquaredB} %
	Suppose we view \eqref{eq:ThyReturnDecompB} as a linear regression of $R_i$ onto the set of regressors $\boldsymbol{R}_{-i}$, and $\mathbf{b}_i$ as the vector of regression coefficients. Assume further: (i) $\mathbf{a}_i^\top \E[\boldsymbol{\Phi}_i ] \neq 0 $; (ii) the idiosyncratic risks are homoskedastic (i.e.\ $\E[\varepsilon_i \varepsilon_j] = 0$ if $i \neq j$ and $= \sigma_\varepsilon^2$ if $i  = j$); and (iii) the variance-covariances $\Var(\boldsymbol{F})$ and $\Var(\boldsymbol{\Phi}_i)$ for all $i$ are positive definite. Then,
	\begin{enumerate}[(a)] 
		\item The regression $\mathsf{R}$-squared decreases (increases) as $\mathbf{a}_i^\top ( \Var( \boldsymbol{\Phi}_i ) - \Var( \boldsymbol{F} ) ) \mathbf{a}_i - \sigma_\varepsilon^2$ becomes more positive (more negative). 

		\item If moreover $\mathbf{a}_i^\top \E[\boldsymbol{\Phi}_i] > 0$, then the regression $\mathsf{R}$-squared is decreasing in $\mathbf{a}_i^\top \E[\boldsymbol{\Phi}_i ]$.  
	\end{enumerate}
\end{cor} 
In the remainder of this paper, we will identify the regression $\Rsquared$ with the \emph{Statistical Arbitrage Risk} (SAR) of an asset $i$. That is, we will say an asset $i$ has a high (low) SAR if it has low (high) $\Rsquared$. 

Corollary~\ref{cor:AssetRSquaredB}(a) provides another way to view the $K$ factor residuals of stock $i$ from \eqref{eq:InsuranceAssetRet}. Condition (ii) is used to simplify the equations and condition (iii) is a mild technical assumption. For the sake of exposition, consider the case when $\sigma_\varepsilon^2$ is negligible, and so the magnitude of the term in Corollary~\ref{cor:AssetRSquaredB}(a) is driven by the positive- or negative-definiteness of the matrix $\Var( \boldsymbol{\Phi}_i ) - \Var( \boldsymbol{F} ) $. The case where this $K \times K$ matrix is \emph{negative-definite} is when the volatility of the factor residuals of asset $i$ is lower than the volatility of the factors themselves. This happens when the factor residuals of asset $i$ do a good job in hedging asset $i$ against its exposure to the $K$ factor risks. Recall from \eqref{eq:InsuranceAssetRet} these $K$ factor residuals are dependent on the factor structure of all the other $N$ risky assets. This implies these $K$ factor residuals can only do a good job in insuring asset $i$ against factor risks if the $N$ other risky assets also highly co-move with asset $i$ itself, which implies a \emph{high} regression $\mathsf{R}$-squared. Given the desirability of these $K$ factor residuals, the holder of asset $i$ will be willing to pay a high price for these $K$ factor residuals, which then pushes \emph{down} their expected returns. This explains why in Corollary~\ref{cor:AssetRSquaredB}(b), there is a negative relationship between the regression $\mathsf{R}$-squared and the SARP. The discussion for the case when that $K \times K$ matrix is positive-definite is analogous. The above discussions are still contingent upon the existence and positivity of such a SARP, which again, is entirely an empirical question we now proceed to answer.

\section{Empirical hypothesis and methodology} 
\label{sec:Methodology}
We summarize the empirical implications of our theoretical discussions.  
\begin{empiricalimp*}
	\hfill
	\begin{enumerate}[(1)]
		\item The expected difference between a given asset's return and some linear combination of other assets' returns can be seen as a \emph{Statistical Arbitrage Risk Premium (SARP)}. Under weak economic and technical conditions, the SARP is non-zero. Moreover, one does \emph{not} need to know a priori what are the underlying factors that drive the economy to compute SARP. 

		\item We can identify regression $\Rsquared$ with \emph{Statistical Arbitrage Risk (SAR)}. We anticipate assets with low SAR (i.e.\ high $\Rsquared$) to have a low SARP, while assets with a high SAR (i.e.\ low $\Rsquared$) to have a high SARP in the cross-section.
	\end{enumerate} 
\end{empiricalimp*} 

The empirical methodology of the paper is separated into two distinct steps. In the first step, we project a given stock's return onto the span of all other stocks' returns to get an empirical approximation of $\mathbf{b}_i$ from \eqref{eq:ThyReturnDecompB}. However, despite the microfoundations of Theorem~\ref{thm:ThyReturnDecomp}, we shall argue it is econometrically non-trivial to execute this projection. We will apply a machine learning method to overcome a critical technical hurdle. In the second step, we will use standard portfolio sort methods from the empirical asset pricing literature to estimate the expectation $\mathbf{a}_i^\top \E[\boldsymbol{\Phi}_i ]$ of \eqref{eq:DiffedExpectations}, which is again the SARP of asset $i$.

\subsection{Data and projection procedure via the elastic-net}
\label{sec:DataEstimation}

\subsubsection{Data sources}
Our data source is standard. We use both the CRSP daily and monthly data from December 31, 1974 to December 31, 2020 with the standard filters. In all of the subsequent empirical analysis, we identify a stock by its \texttt{PERMNO} number as recorded in the CRSP database.
\footnote{
	As is well known with the CRSP and Compustat databases, there are several unique identifiers of equities: \texttt{PERMNO}, \texttt{PERMCO} and \texttt{GVKEY}. Each have their different strengths and weaknesses. We recognize some --- but few --- firms have dual class shares which implies a single firm can have having multiple \texttt{PERMNO}s. Rather than making arbitrary corrections to somehow ``merge'' the time series of returns of these dual class shares, we simply leave the \texttt{PERMNO}s ``as is'' in CRSP. In all, this means our paper focuses on tracking an individual security rather than the individual firm --- although they are synonymous with each other except for those few special cases.
}
Moreover, we only include stocks that, in the past twelve months for any given month end, there are at least 60 days of valid trading returns. This 60 days choice ensures that we do include effectively all stocks, except for the most extremely illiquid or dead stocks, so that our results are not driven only by the liquid and hence most likely large stocks. We also obtain the Fama-French data from Kenneth French's website. 
\begin{figure}
    \centering
%
%
%

\begin{tikzpicture}[scale=0.8]

	\newcommand{\xaxisA}{0}
	\newcommand{\EstTimeLength}{-5.5}
	\newcommand{\AheadTimeLength}{-3}
	\coordinate (BegEstTimeA) at (\xaxisA,0);  
	\coordinate (EndEstTimeA) at ($(BegEstTimeA) + (0,\EstTimeLength)$);  
	\coordinate (AheadTimeA) at ($(EndEstTimeA) + (0,\AheadTimeLength)$);  

	\draw [|-|, thick, >=latex] %
	(BegEstTimeA) node[align=center, anchor=south, xshift=-1ex, color=orange, rotate=90] {\footnotesize{Dec 31,} \\ \footnotesize{1974} } %
		-- %
		(EndEstTimeA) node[align=center, anchor=south, xshift=-1ex, color=orange, rotate=90] { \footnotesize{Month end} \\ \footnotesize{$t - 1=$} \\ \footnotesize{Dec 31, 1975} };  
	\draw [-|, thick, >=latex] %
		(EndEstTimeA) %
		-- %
		(AheadTimeA) node[align=center, anchor=south, xshift=-1ex, color=teal, rotate=90] {\footnotesize{$t =$} \\ \footnotesize{Jan 31, 1976}};  

	\fill [pattern=north east lines, pattern color=orange] ($(BegEstTimeA) + (-0.10,0)$) rectangle ($(EndEstTimeA) + (0.1,0)$);

	\node[anchor=east, align=center, rotate=90, yshift=5ex] at ($(BegEstTimeA) + (0,-0.9)$) %
	{  \footnotesize{\textit{Twelve months} of} \\ \footnotesize{\textit{daily} observations} \\ $ \footnotesize{ d_1,  \ldots, D_{i,t-1} } $ };

	\draw [->, >=latex, color=teal] ($(EndEstTimeA) + (-2,0)$) %
		to[out=250, in=130] %
		node [align=center, anchor=south, xshift=-1.5ex, rotate=90] { \footnotesize{One month} \\ \footnotesize{ahead} } 
		($(AheadTimeA) + (-1.5,0)$); 

	\coordinate (GridTopLeft) at ($(BegEstTimeA) + (2,0)$); 
	\coordinate (GridBottomRight) at ($(EndEstTimeA) + (5,0)$); 

	\fill [pattern=grid, pattern color=orange] ($(GridTopLeft)$) %
		rectangle ($(GridBottomRight)$); %
	\fill [pattern=grid, pattern color=orange] ($(GridTopLeft) + (3.25,0)$) rectangle ($(GridBottomRight) + (6,0)$);

	\node [anchor=south, align=center] at (GridTopLeft) {\footnotesize{1}};
	\node [anchor=south, align=center] at ($(GridTopLeft) + (1.2,0)$) {\footnotesize{$\cdots$}};
	\node [anchor=south, align=center] at ($(GridTopLeft) + (2.5,0)$) {\footnotesize{$i - 1$}};
	\node [anchor=south, align=center] at ($(GridTopLeft) + (3.75,0)$) {\footnotesize{$i + 1$}};
	\node [anchor=south, align=center] at ($(GridTopLeft) + (6,0)$) {\footnotesize{$\cdots$}};
	\node [anchor=south, align=center] at ($(GridTopLeft) + (8.75,0)$) {\footnotesize{$N_{t - 1}$}};

	\fill [pattern=grid, pattern color=black] ($(BegEstTimeA) + (0.75,0)$) %
		node[align=center, anchor=south, xshift=0.5ex] {$i$} %
		rectangle ($(EndEstTimeA) + (0.75,0) + (0.3,0)$) %
		node[align=left, anchor=north, xshift=8ex] { $\boldsymbol{\hat{\beta}}_{i,t-1} \;\text{and}\; \mathsf{R^2}_{i,t-1}$ };

	\node [align=left] at ($(BegEstTimeA) + (2.5,0) + (4.25,-10.5)$) %
	{Return of (\textit{focal}) stock $i$ at month $t$ \\ %
		\quad $R^{\textrm{Foc}}_{i,t} \equiv R_{i,t} = \frac{P_{i,t}}{P_{i,t-1}} - 1$ \\ \\ 
		Return of the \textit{replicate} of stock $i$ at month $t$ \\ %
		\quad $R^{\textrm{Rep}}_{i,t} = (1 - \ind^\top \boldsymbol{\tilde{\beta}}_{i,t-1} ) r_{f,t} + \boldsymbol{R}^\top_{-i,t} \boldsymbol{\tilde{\beta}}_{i,t-1} $, \\  %
		\quad where $\boldsymbol{\tilde{\beta}}_{i,t-1} := \boldsymbol{\hat{\beta}}_{i,t-1} / \smallnorm{\boldsymbol{\hat{\beta}}_{i,t-1}}_1 $ if $\boldsymbol{\hat{\beta}}_{i,t-1} \neq \boldsymbol{0}$, and $:= \boldsymbol{0}$ if otherwise. \\ \\
		Return of longing stock $i$ and shorting its replicate at month $t$ \\ %
		\quad $R^{\textrm{LS}}_{i,t} = R^{\textrm{Foc}}_{i,t} - R^{\textrm{Rep}}_{i,t}$
		};

\end{tikzpicture}

	\caption{\textbf{Projection and replicate construction procedure}. For each month ending at December 31, 1975, January 31, 1976, ..., November 30, 2020, we use the past twelve months' worth of daily observations to project each of stock returns onto the returns span of every other stock. We only consider stocks have at least 60 days worth for daily trading data. That is, for each stock $i = 1, \ldots, N_{t-1}$, suppose $\{ d_{1}, \ldots, D_{i,t-1} \}$ with $D_i \ge 60$ are past twelve months' worth of trading days that end at month $t - 1$. Note that the the number of stocks $N_{t-1}$ in the market at each month end $t-1$ may vary. The returns span for stock $i$ are all those other stocks $j = 1, \ldots, i - 1, i + 1, \ldots, N_{t-1}$ that have trading days that at least overlap with that of stock $i$, so $\{ d_{1}, \ldots, D_{i,t-1} \} \subseteq \{ d_{1}, \ldots, D_{j,t-1} \}$. 
    }
    \label{fig:fig_estimation_procedure}
\end{figure}

\begin{rem}[Filters and sampling period]
	We subset only for US common equities (i.e.\ \texttt{SHRCD} code of 10 or 11), and only those that are listed in the NYSE, AMEX or NASDAQ (i.e.\ \texttt{EXCHCD} code of 1, 2 or 3). In addition, we start our data sample from 1974 because the CRSP datasets only started to include NASDAQ stocks in December 1972. We start in December 1974 to allow an additional year of buffering for good measure. Effectively this means the number of stocks in CRSP pre-1974 and post-1974 are structurally different in several magnitudes; see \cite[\S 7.1.2]{bali2016empirical} for a detailed discussion. Our estimation and projection method is clearly sensitive to total number of stocks. Starting our analysis pre-1974 might bias our results simply due to CRSP data limitations. 
\end{rem}

\subsubsection{Projection}
\label{sec:ProjectionProcedure}
The first step in the empirical test of Theorem~\ref{thm:ThyReturnDecomp} is to project a given stock's returns onto the returns span of all other stocks. Our projection procedure is summarized in Figure~\ref{fig:fig_estimation_procedure}. For months ending $t - 1 = $ December 31, 1975, January 31, 1976, ..., November 30, 2020, we use the past twelve months' worth of daily observations to project each of stock $i$'s returns onto the returns of \textit{all other} stocks. We reserve a one month gap for returns realization in a procedure to be described in Section~\ref{sec:PortfolioSort}. Unless specified otherwise, we will denote $t - 1$ as the end of the projection month, and $t$ as the one-month ahead returns realization date; that is, we set $t = $ January 31, 1976, February 29, 1976, \ldots, December 31, 2020. Figure~\ref{fig:severalthings}(a) shows the number of stocks that have at least 60 past trading days in for each given month $t - 1$. Observe since we start our projection procedure on December 31, 1975, we require daily data starting from December 31, 1974. 

Let $N_{t - 1}$ be the total number of stocks traded in the market at month $t - 1$. The return vector of stock $i$ at month $t - 1$ and the returns span of all other assets are, respectively: 
\begin{subequations}
\begin{gather}
	y_{i,t-1} = 
	\begin{pmatrix}
		R_{i, d_{1}} \\ 
		\vdots \\
		R_{i, d_{D_{i,t-1}}} 
	\end{pmatrix}_{D_{i,t-1} \times 1} \label{eq:LHSRHSVariablesL} 
	\\
	\mathbf{X}_{i,t-1} =
	\begin{pmatrix}
		R_{1, d_{1}} & \ldots & R_{i-1, d_{1}} & R_{i+1, d_{1}} & \ldots & R_{N_{t-1}, d_{1}} \\ 
		\vdots  &   & \vdots & \vdots &  & \vdots \\ 
		R_{1, d_{D_{i,t-1}}} & \ldots & R_{i-1, d_{D_{i,t-1}}} & R_{i+1, d_{D_{i,t-1}}}  & \ldots & R_{ N_{t-1} , d_{D_{i,t-1}}}  
	\end{pmatrix}_{ D_{i,t-1} \times (N_{t-1} - 1)  } \label{eq:LHSRHSVariablesR}
\end{gather}
\label{eq:LHSRHSVariables}
\end{subequations}
where $R_{i,d}$ is the daily return of stock $i$, $D_{i,t-1}$ is the number of trading days of stock $i$ in the past 12 months ending at month $t -1$. The dimensions of \eqref{eq:LHSRHSVariablesR} are approximately $250 \times 5000$ for each stock $i$. There are $T = 539$ number of months from December 31, 1975 to November 30, 2020. Thus we run a total of approximately $T \times 5000 \approx 2.7$ million projections in this paper. 

In this paper we will use the \emph{elastic-net estimator} developed by \cite{Zou2005} to empirically project a given stock's return onto the returns span of all other stocks. We defer a detailed and technical discussion of the elastic-net in our context to Section~\ref{sec:EstimationDiscussions}. But if we are interested in estimating \emph{linear} relationships as in Theorem~\ref{thm:ThyReturnDecomp}, why do we not use the workhorse \emph{ordinary least squares} (OLS) estimator? The design matrix \eqref{eq:LHSRHSVariablesR} of returns to evaluate our empirical hypothesis is necessarily a $T \times N$ matrix, where $T \approx 250$ is the number of days, and $N \approx 5000$ is the total number of traded stocks. This is a case where $T \ll N$. This means the $N \times N$ matrix $\mathbf{X}_{i,t-1}^\top \mathbf{X}_{i,t-1}$ is \emph{not} full rank. The OLS estimator is thus necessarily \emph{not} well defined. In contrast, the elastic-net is a machine learning method that explicitly allows for ``wide'' $T \ll N$ regressors.

We denote the elastic-net projection coefficient vector of stock $i$ at month $t-1$ as $\boldsymbol{\hat{\beta}}_{i, t-1} \in \R^{N_{i,t-1}}$, and the resulting \textit{coefficient of determination (``R-squared'')} as $\Rsquared_{i,t-1}$. Note and recall the conventional definition of $\Rsquared$ is, 
\begin{equation*}
	\Rsquared_{i,t-1} := 1 - \frac{\sum_{s=1}^{D_{i, t-1}} ( R_{i,d_s} - \hat{y}_{i,t-1} )^2 }{ \sum_{s=1}^{D_{i, t-1}} (R_{i,d_s} - \hat{\mu}_{i,t-1} )^2  }, 
\end{equation*}
where $\hat{y}_{i,t-1} := \mathbf{X}_{i,t-1}\boldsymbol{\hat{\beta}}_{i,t-1}$ is the fitted value, and $\hat{\mu}_{i,t-1}$ is the sample mean; they are explicitly given by  
\begin{equation*}
	\hat{y}_{i,t-1} = \mathbf{X}_{i,t-1}\boldsymbol{\hat{\beta}}_{i,t-1}, \quad 
	\hat{\mu}_{i,t-1} = \frac{1}{D_{i,t-1}} \sum_{s = 1}^{D_{i,t-1}} R_{i,d_s}.  
\end{equation*}

We emphasize we are only using the elastic-net as a projection method for constructing the replicates, and we do \emph{not} use it for statistical inference. The statistical inference claims are on the expected returns of SARP, which we will discuss beginning in Section~\ref{sec:PortfolioSort}. As a result, despite the large number of projections that we run at this stage, we do not suffer from the multiple hypothesis testing problem that has been discussed in the recent empirical asset pricing literature by \cite{Harvey2016}. Other than the use in calculating $\Rsquared$, we do \emph{not} use the fitted value $\hat{y}_{i, t - 1}$ in any other subsequent steps. This is in contrast to the recent literature on financial applications of machine learning (say \cite{Gu2018} and others) where they use the fitted value (or predicted value when one uses $\mathbf{X}_{i, t}$ instead of $\mathbf{X}_{i, t - 1}$) in assessing model forecasting accuracy.

\begin{rem}[Not projecting onto the intercept]
	\label{rem:NoIntercept}
	We deliberately do \emph{not} include an intercept as a regressor in \eqref{eq:LHSRHSVariablesR}. Including an intercept together with the sparseness property of the elastic-net estimator will attribute a stock with stale prices with extremely high $\Rsquared$. See Section~\ref{sec:InterceptEstimation} for a more detailed discussion. Other than the projection method here and in Section~\ref{sec:ComparingAgainstFF}, unless noted otherwise, all subsequent statistical inference tests that employ a linear regression will include an intercept.
\end{rem}

\begin{rem}[Why elastic-net and not other machine learning methods?]
	Out of a myriad of machine learning methods, why did we choose the elastic-net estimator to test our empirical implication? Simply put, we regard the elastic-net estimator as a parsimonious method that allows us to test our theoretical prediction. We defer to Section~\ref{sec:EstimationDiscussions} for detailed technical discussions of why we particularly use and prefer the elastic-net estimator in this paper.
\end{rem}

\begin{rem}[Overlapping data]
	It is evident we are using overlapping time data. Overlapping returns data in traditional empirical asset pricing raises several technical inference issues revolving around autocorrelations (see \cite{Hansen1980} and a recent discussion by \cite{Hedegaard2016}). However, the ``wide'' regressors in our setting imply we actually gain $\approx 5000$ new cross-sectional data points for each month advance. Because of the sheer amount of new entering cross sectional data, it is not necessarily true that the projected coefficient ending at months $t - 2$ and $t - 1$ would be quantitatively similar even if they only differ by a one month step. 
\end{rem}

\subsection{Replicate construction}
\label{sec:PortfolioConstruction}
For each month end $t - 1 $ and each stock $i = 1, \ldots, N_{t - 1}$, we collect the projected coefficient $\boldsymbol{\hat{\beta}}_{i,t-1}$ and the $\Rsquared_{i,t-1}$. We track stock $i$'s \emph{one-month ahead} return $R_{i,t}$ from $t - 1$ to $t$. For subsequent exposition clarity, we will sometimes call stock $i$ as the \emph{focal stock} and write it's return at month $t$ as,  
\begin{equation}
	R_{i,t}^{\textrm{Foc}} \equiv R_{i,t}. 
	\label{eq:Actual} 
\end{equation}

Next, we introduce the \emph{replicate} (portfolio) of a stock $i$. This is a key idea of this paper. Let $\boldsymbol{R}_{-i,t} := [ R_{1,t}, \ldots, R_{i-1,t}, R_{i+1,t}, \ldots, R_{N_{t-1}, t} ]^\top$ be the vector of month $t$ returns for all stocks except stock $i$. We wish to treat the projected coefficients $\boldsymbol{\hat{\beta}}_{i, t - 1}$ as investment weights into each of the $N_{t-1} - 1$ number of stocks. However, the regularization nature of the elastic-net causes the entries of these estimated coefficients to be small in magnitude. So if we were to directly use $\boldsymbol{\hat{\beta}}_{i, t - 1}$ as investment weights, the result would be a very small allocation into risky component of the portfolio. To have a more reasonable risky component, we normalize $\boldsymbol{\hat{\beta}}_{i, t - 1}$ by its $L^1$ norm and write,
\footnote{
	For any vector $\mathbf{x} = (x_1, \ldots, x_n) \in \R^n$, we define its $L^1$ norm as $\norm{\mathbf{x}}_1 := \sum_{j = 1}^n \abs{x_j}$. Note that the $L^1$ norm shows up explicitly in the objective function of the elastic-net estimator. The $L^2$ norm also shows up in the elastic-net objective function; as is usual, $\norm{\mathbf{x}}_2 := \sqrt{\sum_{j = 1}^n x_j^2}$. However, we choose to normalize by $L^1$ and not $L^2$ because of scaling. We have inequality $\norm{\mathbf{x}}_2 \le \norm{\mathbf{x}}_1$. Economically and empirically, this implies scaling by the $L^2$ norm will result in a highly levered equity component for the replicate for stock $i$. The $L^1$ scaling generates more a more moderate result.
}
\begin{align}
	\boldsymbol{\tilde{\beta}}_{i, t - 1}
	:=
	\begin{dcases}
		\boldsymbol{0} & \text{if $\boldsymbol{\hat{\beta}}_{i, t - 1} = \boldsymbol{0}$,} \\
		\frac{\boldsymbol{\hat{\beta}}_{i, t - 1}}{\smallnorm{\boldsymbol{\hat{\beta}}_{i, t - 1}}_1} & \text{otherwise.} 
	\end{dcases}
	\label{eq:WeightNormalization}
\end{align}
Since investment weights must sum to one, we place the remainder of the weights $\ind^\top \boldsymbol{\tilde{\beta}}_{i,t-1}$ into the risk free asset with returns $r_{f,t}$, and here $\ind$ is a vector of ones of conformable dimensions. In all, the month $t$ return of stock $i$'s \textit{replicate} is, 
\begin{equation}
	R_{i,t}^{\textrm{Rep}} := (1 - \ind^\top \boldsymbol{\tilde{\beta}}_{i,t-1} ) r_{f,t} + \boldsymbol{R}_{-i,t}^\top \boldsymbol{\tilde{\beta}}_{i,t-1}. 
	\label{eq:Ghost} 
\end{equation}

\begin{figure}[htp]
    \centering
    \includegraphics[scale=0.6]{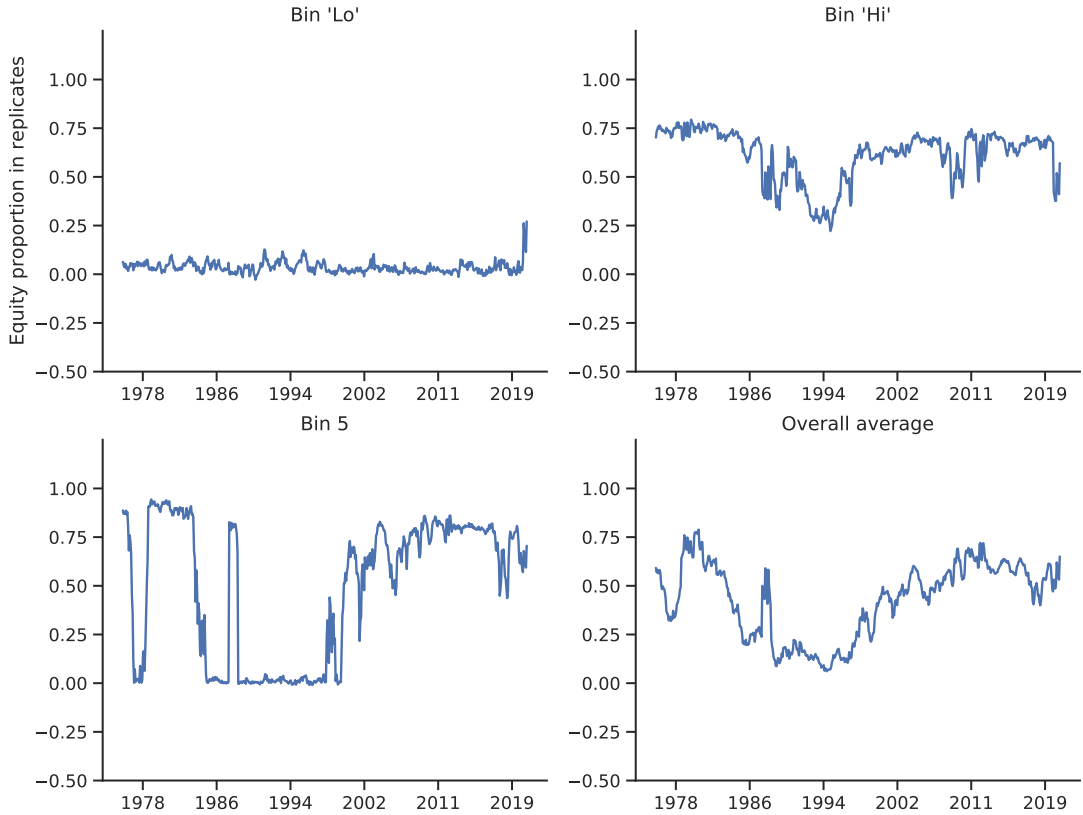}
    \caption{ \textbf{Proportion of equity invested in the replicates. } %
			We consider the average amount of equity-only components that are invested by each replicate portfolio. Stocks are sorted into their method $m$ $\Rsquared$ decile bins $k = \textrm{'Lo'}, \ldots, \textrm{'Hi'}$. At the end of the estimation month $t - 1$, we have the normalized vector $\boldsymbol{\tilde{\beta}}_{i, t-1}$ of asset $i$. The quantity $\ind^\top \boldsymbol{\tilde{\beta}}_{i, t-1}$ is the proportion allocated to the equity-only components of stock $i$'s replicate as from \eqref{eq:Ghost}. The average equity-only proportion across all the replicates in bin $k$ at month $t - 1$ is the quantity $\frac{1}{| B_{t-1}^k |} \sum_{i \in B_{t-1}^k } \ind^\top \boldsymbol{\tilde{\beta}}_{i,t-1}$. Here we plot time series of this average equity-only proportion for the bins $k = \textrm{'Lo'}, 5$ and $\textrm{'Hi'}$, and the $y$-axis are expressed in decimals (e.g.\ 0.10 means 10\%). The plot labeled ``Overall average'' plots the overall average of these equity-only components across all bins, which is the quantity $\frac{1}{10} \frac{1}{| B_{t-1}^k |}  \sum_{k=1}^{10} \sum_{i \in B_{t-1}^k } \ind^\top \boldsymbol{\tilde{\beta}}_{i,t-1} $. The sampling period is from December 1975 to November 2020. 
}
    \label{fig:cs_equity_in_ghosts} 
\end{figure} 

Finally, we track the \textit{long-short} return of stock $i$ against its replicate,  
\begin{equation}
	R_{i,t}^{\textrm{LS}} := R_{i,t}^{\textrm{Foc}} - R_{i,t}^{\textrm{Rep}}. 
	\label{eq:ActualMinusGhost} 
\end{equation}
This long-short return \eqref{eq:ActualMinusGhost} will proxy for the difference $R_i - \mathbf{b}_i^\top \boldsymbol{R}_{-i}$ in Theorem~\ref{thm:ThyReturnDecomp}, and is the key emphasis of study in this paper. We will use conventional portfolio sort methods of the empirical asset pricing literature to estimate the expectation $\E[R_i] - \mathbf{b}_i^\top \E[\boldsymbol{R}_{-i}]$, which then is equal to the theoretically motivated SARP $\mathbf{a}_i^\top \E[\boldsymbol{\Phi}_i ]$ of stock $i$. 

\begin{rem}[Projection method for constructing mimicking portfolios]
	\cite{Breeden1989} and \cite{Lamont2001} use analogous methods to \eqref{eq:Ghost} to construct a mimicking factor out of tradable base assets. \cite{2006_JF_Ang} use also an analogous method to construct a factor mimicking aggregate volatility risk. While our approach in \eqref{eq:Ghost} seems identical to these previous methods, we stress the dimensionality of the regressors is substantially different. The number of base assets in those aforementioned methods is small, usually in the range of five to ten. In contrast, our base assets are effectively every other stock other than the stock $i$ itself, which number in the thousands.\cite{wurgler2002does}
	\footnote{
		We are grateful for an anonymous referee for pointing out this useful reference to us.
	}
	uses an idea that is conceptually similar --- but operationally quite different --- to this paper in evaluating arbitrage risk. \cite{wurgler2002does} first use a sorting method to closely match a stock $i$ with three other stocks that is most closest to it on size and book-to-market, and then use a linear regression on of the returns of stock $i$ onto these three stocks. Similar to us, they use the regression coefficients to construct a replicate of stock $i$. However, the selection procedure of \cite{wurgler2002does} by portfolio sort is not entirely statistical in nature. Indeed, a priori there's no reason why the time series statistical properties of any given stock $i$ would be best matched by the top three stock matched on size and value, even if size and value are well known priced factors. Moreover, the focus of \cite{wurgler2002does} is to study ``demand curves'' for stocks and to identify this effect, the authors only constrain their study to 259 stocks addition into the S\&P 500 index over a 13 year period. In contrast, the focus of our study is SARP and we consider effectively all stocks (so not just those in the S\&P 500) at any given point in time and over a 45 year period.
\end{rem}

\subsection{Portfolio formation and sort}
\label{sec:PortfolioSort} 
Having defined three types of returns \eqref{eq:Actual}, \eqref{eq:Ghost} and \eqref{eq:ActualMinusGhost} associated with stock $i$, we now proceed to construct portfolios. We use our theoretical discussions as guidance: Corollary~\ref{cor:AssetRSquaredB} anticipates we should find a negative relationship between the regression $\mathsf{R}$-squared (the SAR of stock $i$) and the SARP of stock $i$. At the end of the month $t - 1$, we sort each stock $i$ by its $\Rsquared_{i,t-1}$ into \textit{decile} bins. Let $B_{t-1}^k$ be the set of stocks in the $k$-th bin at month $t - 1$. We organize the bins in ascending order, so bin $k = 1$ (labeled ``Lo'') consists of stocks with the lowest $\Rsquared$'s, and bin $k = 10$ (labeled ``Hi'') consists of stocks with the highest $\Rsquared$'s. 

We focus on equal-weighted portfolios in this paper (see Remark~\ref{rem:ValueWeights} for a discussion of the issues in evaluating value-weighted SARP). The equal-weighted \textit{excess returns} of bin $k = \textrm{Lo}, 2, \ldots, \textrm{Hi}$ of the stocks is standard:  
\begin{equation}
	\bar{R}^{k}_{\textrm{Foc}, t}
	= \frac{1}{ \abs{ B_{t-1}^k} } \sum_{i \in B_{t-1}^k } ( R_{i,t} - r_{f,t} ), \\ 
	\label{eq:PortfolioActual}
\end{equation}
where we denote $\abs{B}$ as the cardinality of a set $B$. In this paper, we will view and assume the replicate of stock $i$ to have the same equal-weight as its focal counterpart,
\begin{equation}
	\bar{R}^{k}_{\textrm{Rep}, t}
	:= \frac{1}{ \abs{ B_{t-1}^k} } \sum_{i \in B_{t-1}^k } (
	R_{i,t}^{\textrm{Rep}} - r_{f,t} ).
	\label{eq:PortfolioGhost}
\end{equation}

Finally, the construction of the equal-weighted portfolio of the long-short of the focal stocks versus their replicates is now immediate thanks to the aforementioned equivalent weighting assumption. We define the equal-weighted portfolio return of the \textit{long-focal, short-replicate position} in bin $k$ as, 
\begin{equation}
	\bar{R}^k_{\textrm{LS}, t}
	:= \bar{R}^k_{ \textrm{Foc}, t} - \bar{R}^k_{ \textrm{Rep}, t } = \frac{1}{\abs{ B_{t-1}^k }} \sum_{i \in B^k_{t-1} } R_{i,t}^\textrm{LS}.
	\label{eq:PortfolioLS}
\end{equation}
Note \eqref{eq:PortfolioLS} does not subtract a risk-free rate as it is the difference of two excess returns.

\begin{rem}[Ambiguity in value-weighted SARP]
	\label{rem:ValueWeights}
	Value-weighting causes several causes for concern in evaluating SARP. Value-weights for focal stocks in each bin $k$ is conventional. The cause for caution is what should one assume for the value-weights of the replicates. Perhaps the most natural way is just assume the replicate of stock $i$ has the same value-weight as stock $i$ itself. While this value-weight assumption for the replicates might seem natural, it is also ambiguous. For stocks with high SAR there is no contention: for these high SAR stocks, their replicates are effectively just the risk-free asset, and so using value-weights in this way would just produce the equity risk premium. But for stocks with low SAR, there will be many similar peer stocks in their replicates. It is possible that a focal stock has large size, but the elastic-net statistically matches its peers with small stocks, and vice-versa. Thus the total value of the replicate portfolio of stock $i$ could be substantially different than that of the focal stock $i$ itself. Assigning the replicate of stock $i$ to have the same value-weight of the focal stock $i$ in bin $k$ may or may not reflect the size of the latter. Indeed, the same concern also holds in converse. This ambiguity in what ``value-weighting'' means for the replicates is what drives our paper to focus on the most parsimonious weighting scheme --- equal-weights. 
\end{rem}

\section{Empirical results}  
\label{sec:MainResults}
We can now translate our theoretical predictions of Section~\ref{sec:TheoreticalMotivation} to concrete quantities for an empirical investigation: (i) By Corollary~\ref{cor:NonZeroAssetInsPrem}, SARP should be non-zero (and indeed positive). That is, we should expect the time-series average of $\bar{R}^k_{\textrm{LS}, t}$ of \eqref{eq:ActualMinusGhost} to be non-zero in the data; and (ii) More importantly by Corollary~\ref{cor:AssetRSquaredB}, SARP should be increasing with SAR. In other words, stocks that have high SAR (low $\Rsquared$) should have high SARP, and stocks with low SAR (high $\Rsquared$) should have low SARP.

\subsection{Summary statistics}
\label{sec:SummaryStatistics}

\subsubsection{Characteristics summary statistics}
\label{sec:CharacteristicsStatistics}
Let's first investigate the average characteristics of stocks that stocks that are decile sorted by elastic-net $\Rsquared$'s. Let $Z_{i,t - 1}$ be some scalar characteristic of stock $i$ at month $t - 1$. For each bin $k$, we compute the simple average of stocks' characteristics $\bar{Z}_{t - 1}^k := \frac{1}{\abs{B_{t - 1}^k}} \sum_{i \in B_{t - 1}^k} Z_{i, t -1}$. The numbers in Table~\ref{tab:consolidated_characteristics_results} show the time-series mean, standard deviation, $5^{\mathrm{th}}$ and $95^{\mathrm{th}}$ percentiles of these characteristics $\bar{Z}_{t - 1}^k$ for each bin $k$.

Result (i) of Table~\ref{tab:consolidated_characteristics_results} shows the elastic-net $\Rsquared$ characteristic $Z_{i, t - 1} = \Rsquared_{i, t - 1}$. The elastic-net is capturing a wide range of projection $\Rsquared$'s from effectively 0\% in the lowest bin to 62\% in the highest bin.
\footnote{ 
While there is some resemblance, we caution that the positive association of $\Rsquared$ and number of non-zero elements in the estimated elastic-net coefficient vector is different from the OLS case where including more regressors necessarily and mechanically raises $\Rsquared$. In our elastic-net application, the number of regressors $N_{t-1}$ remains \textit{fixed}, and it is an estimation outcome that only few of them have non-zero coefficient loadings. 
}
As a matter of comparison and for subsequent robustness discussions in Section~\ref{sec:ComparingAgainstFF}, at the end of month $t - 1$ we take each stock $i$'s past twelve months' of daily returns and project them onto the daily returns of the \cite{Fama2015} five factors using OLS. We collect the resulting FF5 OLS $\Rsquared$ and this is also a characteristic for each stock $i$ for month $t - 1$. Result (ii) shows the summary statistics of the FF5 OLS $\Rsquared$ characteristic for stocks sorted by elastic-net $\Rsquared$. It is quite interesting to observe that stocks with high SAR (low elastic-net $\Rsquared$) have also low corresponding FF5 OLS $\Rsquared$, and likewise stocks with low SAR (high elastic-net $\Rsquared$) have high FF5 OLS $\Rsquared$.

Results (iii) and (iv) examine the quality of our replicate construction procedure. Recall \eqref{eq:Ghost}. The characteristic $Z_{i, t - 1} = \textrm{\textit{Number of non-zero entries in $\boldsymbol{\hat{\beta}}_{i, t - 1}$}}$ is the number of risky peers used to construct the replicate of stock $i$ and (iii) shows this result. Note the sparseness of the elastic-net projection vector: while $N_{t - 1}$ ranges in the thousands, the number of non-zero entries in the projection vector is remarkably low. For stocks with the highest SAR, it is essentially so ``unique'' that the replicate of this stock is just the risk-free asset. In other words, the next best alternative of not investing into a truly unique stock is simply the risk-free asset. In contrast for stocks with the lowest SAR, one can statistically identify about 82.58 risky peers. So even if one forgoes investing into a ``ubiquitous'' stock, one can construct its replicate consisting of 82.58 peer stocks that have similar statistical properties. Next, the characteristic $Z_{i, t - 1} = \ind^\top \boldsymbol{\tilde{\beta}}_{i, t - 1}$ is the proportion of equity that is used to construct the replicate of stock $i$. Result (iv) shows we allocate only 4\% into equity components (so 96\% into the risk-free asset) of the replicates for stocks with the highest SAR, but we allocate on average 60\% into equity components of the replicates for stocks with the lowest SAR. At least with our elastic-net projection procedure and even for stocks with the highest SAR, we basically never achieve an 100\% allocation into equity for the replicates of any stock; an 100\% allocation can happen if there is a stock $l$ physically different from stock $i$ but are otherwise statistically equivalent. That is to say, \emph{all} stocks are effectively unique but there's still nonetheless a wide range of heterogeneity.

Results (v) to (vii) show the market capitalization, book-to-market and dollar volume liquidity characteristics. From result (v), low SAR stocks tend to be smaller stocks while high SAR stocks tend to be bigger stocks. Nonetheless, we should keep in mind the magnitudes and also within decile bin heterogeneity. For low SAR stocks the average market capitalization is about \$1.2 billion with a standard deviation of \$9.3 billion, and for the high SAR stocks the average is \$6.1 billion with a standard deviation of \$26.2 billion. In all, while the high SAR bin holds on average ``mega-sized'' stocks, the stocks in the low SAR bin are not exclusively microcap stocks either. Hence our subsequent cross-sectional results are unlikely to be driven by a size effect. Result (vi) shows an interesting value tilt for stocks with high SAR and a growth tilt for stocks with low SAR. Finally, result (vi) shows the dollar volume liquidity (VOLD) characteristic. In particular, $\textrm{VOLD}_{i,t - 1}$ is defined as the product of trading volume of stock $i$ on the last day of month $t - 1$ and the closing price of stock $i$ on the last day of month $t - 1$, then all divided by 1,000,000. We see that on average, high SAR stocks have a lower liquidity than low SAR stocks. On the surface, results (v) to (vii), it appears size, book-to-market and liquidity are related to a stock's SAR. We will explicitly control for these three characteristics, among others, in Section~\ref{sec:BivariateSorts} to ensure our results are not driven by the well-known pricing qualities of these characteristics in the cross-section.

\begin{figure}[htp]
	\centering
	\begin{minipage}[h]{0.55\textwidth} 
		\centering
			\begin{subfigure}{1\textwidth}
				\centering
				\includegraphics[scale=0.35]{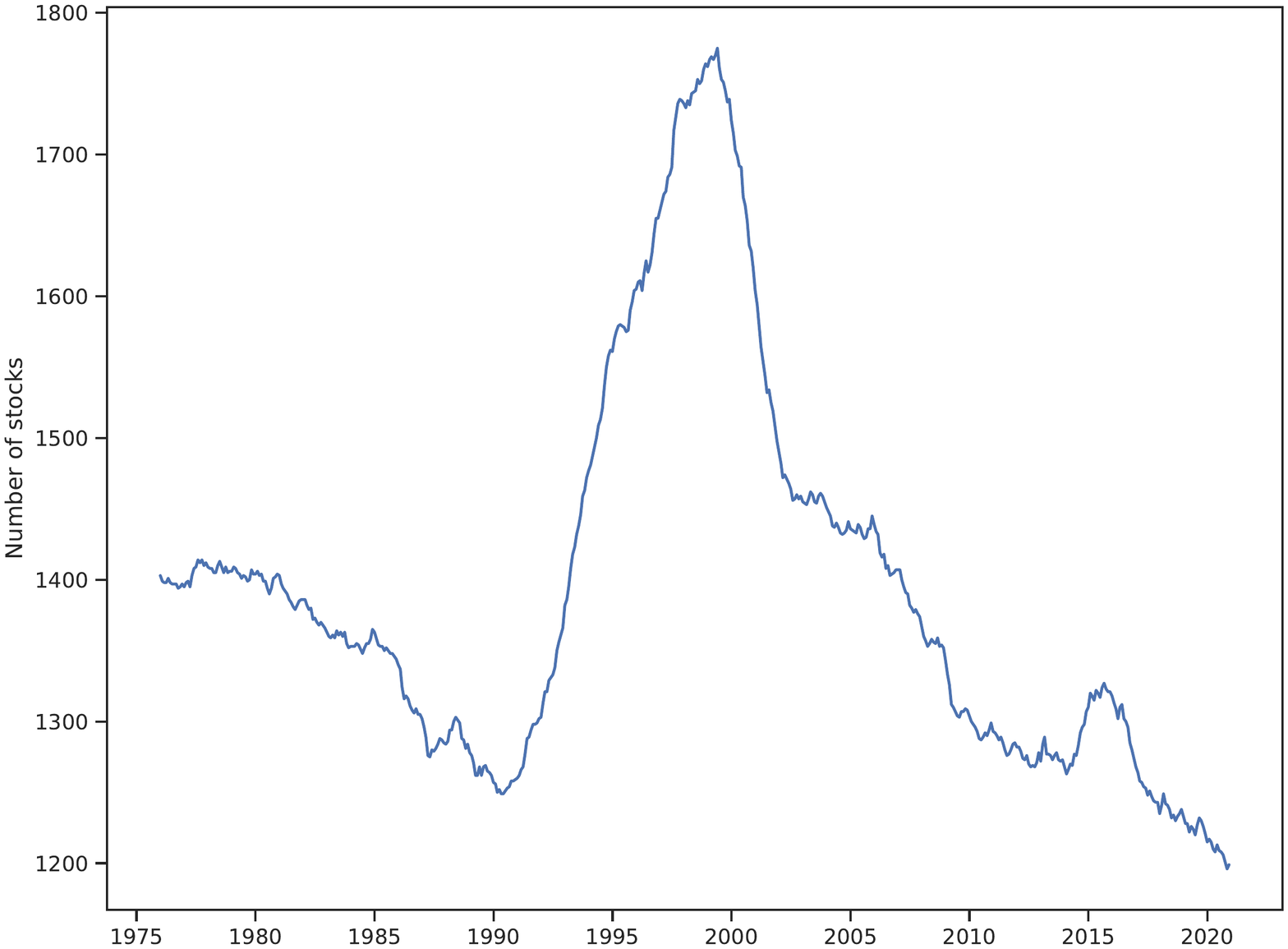}
				\subcaption{Dimensionality value $N_{t - 1}$}
			\end{subfigure} 
		\\
			\begin{subfigure}{1\textwidth}
				\centering
				\includegraphics[scale=0.35]{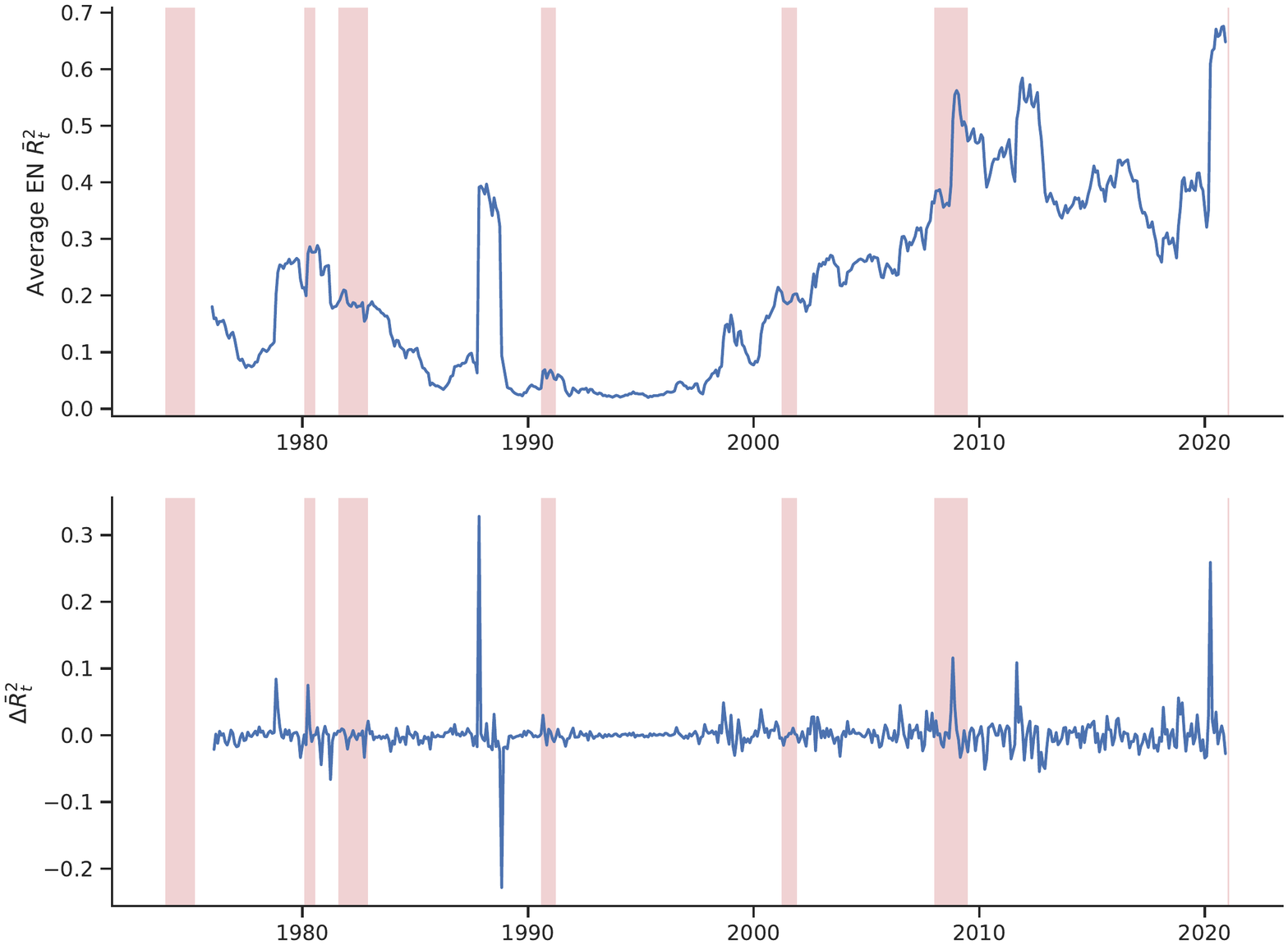}
				\subcaption{Average EN $\Rsquared$ and lagged its difference}
			\end{subfigure} 
	\end{minipage}
	\hfill
	\begin{minipage}[ht]{0.40\linewidth} 
			
\begin{tabular}{@{\extracolsep{5pt}}lccc} 
\\[-1.8ex]\hline 
\hline \\[-1.8ex] 
\\[-1.8ex] & (1) & (2) & (3)\\ 
\\[-1.8ex] & \multicolumn{3}{c}{EN $\Delta \bar{\mathsf{R}}^2_{t}$} \\ 
\hline \\[-1.8ex] 
 $\Delta\log\textrm{INDPRO}_{t}$ & 0.092 & 0.305 & 0.229 \\ 
  & (0.278) & (0.951) & (0.871) \\ 
  & & & \\ 
 $\Delta\log\textrm{PCE}_{t}$ & $-$0.820 & $-$1.082 & $-$0.893 \\ 
  & ($-$2.138) & ($-$2.074) & ($-$2.044) \\ 
  & & & \\ 
 $\Delta\textrm{UNRATE}_{t}$ & $-$0.750 & $-$0.980 & $-$0.505 \\ 
  & ($-$0.879) & ($-$1.118) & ($-$0.694) \\ 
  & & & \\ 
 $\Delta\log\textrm{UMCSENT}_{t}$ & $-$0.081 & $-$0.075 & $-$0.047 \\ 
  & ($-$3.079) & ($-$3.294) & ($-$2.262) \\ 
  & & & \\ 
 $\Delta\log\textrm{INDPRO}_{t-1}$ &  & $-$0.289 & $-$0.111 \\ 
  &  & ($-$1.195) & ($-$0.587) \\ 
  & & & \\ 
 $\Delta\log\textrm{PCE}_{t-1}$ &  & $-$0.046 & 0.075 \\ 
  &  & ($-$0.148) & (0.270) \\ 
  & & & \\ 
 $\Delta\textrm{UNRATE}_{t-1}$ &  & 0.110 & 0.472 \\ 
  &  & (0.224) & (1.008) \\ 
  & & & \\ 
 $\Delta\log\textrm{UMCSENT}_{t-1}$ &  & $-$0.031 & $-$0.029 \\ 
  &  & ($-$1.617) & ($-$1.592) \\ 
  & & & \\ 
 $\textrm{MktRF}_{t}$ &  &  & $-$0.183 \\ 
  &  &  & ($-$2.746) \\ 
  & & & \\ 
 Constant & 0.002 & 0.003 & 0.004 \\ 
  & (1.431) & (1.519) & (1.918) \\ 
  & & & \\ 
\hline \\[-1.8ex] 
Observations & 514 & 513 & 513 \\ 
R$^{2}$ & 0.073 & 0.089 & 0.176 \\ 
Adjusted R$^{2}$ & 0.065 & 0.075 & 0.161 \\ 
\hline 
\hline \\[-1.8ex] 
\end{tabular} 

			\subcaption{Regressing the difference of EN $\Rsquared$ on macro variables} 
	\end{minipage} %
	\caption{\textbf{Number of regressors, average elastic-net $\Rsquared$, and macro regressions}. %
		Figure (a) shows the time series of the total number of traded stocks $N_{t - 1}$ in the US at any given point in time. The elastic-net $\Rsquared$ for each stock $i$ at month $t - 1$ is denoted as $\Rsquared_{i, t - 1}$. The top panel of Figure (b) plots the time series of the averaged $\Rsquared$ across all stocks $\bar{\Rsquared}_t$, and the bottom panel plots the difference $\Delta \bar{\Rsquared}_t := \Rsquared_t - \Rsquared_{t - 1}$. Table (c) regresses the time series $\Delta \bar{\Rsquared}_t$ onto the following regressors' difference and its lag: industrial production, personal consumption expenditure, unemployment rate, University of Michigan consumer sentiment, and the market factor. All macro variables data are from FRED Economic Data of the St.\ Louis Federal Reserve. Since UMCSENT is only readily available from January 1978, the regressions' sampling period is monthly from January 31, 1978 to December 31, 2020.
	}
	\label{fig:severalthings}
\end{figure}

Pursuant to Table~\ref{tab:consolidated_characteristics_results}(i), we directly plot the time series of the cross-sectionally averaged $\Rsquared$ characteristic in Figure~\ref{fig:severalthings}(b). Again, as motivated by our theoretical discussions, we regard assets with low $\Rsquared$ to have high SAR, and assets with high $\Rsquared$ to have low SAR. As a consequence, the plots in Figure~\ref{fig:severalthings}(b) can thus be regarded as an ``average'' SAR of the economy at any given point in time. By inspection, it appears the average SAR of the economy is countercyclical and spikes considerably during times of financial distress (e.g.\ 1980-1991 US savings and loan crisis, 1998 Russian financial, 2008 - 2009 Great Recession, Greek government debt crisis 2011-2012, March 2020 COVID-19 crash, and others). These plots strongly suggest SAR (and the resulting SARP of assets) is not simply a statistical construct but is also capturing systematic shocks. We formally test the association of SAR with macroeconomic variables in Figure~\ref{fig:severalthings}(c). The regression results corroborate the visual inspection of Figure~\ref{fig:severalthings}(b): shocks to the average SAR is countercyclical, in that it is negatively associated with shocks to both personal consumption expenditure and consumer sentiment.

\begin{landscape}
\begin{table}[h]
    \centering
		\begin{tabular}{lcccccccccc}
\toprule
                         \phantom{Characteristic} & EN $\mathsf{R^2}$ Lo &                2 &                3 &                4 &                5 &                6 &                7 &                8 &                 9 & EN $\mathsf{R^2}$ Hi \\
\midrule
                   (i) \textbf{EN $\mathsf{R^2}$} &                -0.01 &            -0.00 &             0.02 &             0.07 &             0.12 &             0.18 &             0.25 &             0.33 &              0.42 &                 0.62 \\
                                                  &               [0.01] &           [0.02] &           [0.07] &           [0.13] &           [0.18] &           [0.22] &           [0.25] &           [0.28] &            [0.29] &               [0.28] \\
                                                  &       (-0.03, -0.00) &    (-0.01, 0.01) &    (-0.00, 0.14) &    (-0.00, 0.37) &    (-0.00, 0.52) &    (-0.00, 0.62) &     (0.00, 0.71) &     (0.01, 0.79) &      (0.03, 0.86) &         (0.11, 0.98) \\
             (ii) \textbf{FF5 OLS $\mathsf{R^2}$} &                 0.08 &             0.07 &             0.07 &             0.09 &             0.11 &             0.14 &             0.17 &             0.21 &              0.25 &                 0.31 \\
                                                  &               [0.07] &           [0.07] &           [0.07] &           [0.09] &           [0.11] &           [0.13] &           [0.15] &           [0.17] &            [0.19] &               [0.21] \\
                                                  &         (0.01, 0.22) &     (0.01, 0.20) &     (0.01, 0.21) &     (0.01, 0.27) &     (0.01, 0.35) &     (0.01, 0.42) &     (0.01, 0.48) &     (0.01, 0.54) &      (0.02, 0.60) &         (0.02, 0.68) \\
(iii) \textbf{Non-zero entries of EN proj vector} &                 0.50 &             0.72 &             2.34 &             6.17 &            11.37 &            17.76 &            25.22 &            34.07 &             46.85 &                82.58 \\
                                                  &               [0.60] &           [1.63] &           [5.73] &          [11.66] &          [17.00] &          [21.89] &          [26.84] &          [32.28] &           [38.45] &              [54.92] \\
                                                  &         (0.00, 1.00) &     (0.00, 2.00) &    (0.00, 12.00) &    (0.00, 32.00) &    (0.00, 46.00) &    (0.00, 59.00) &    (1.00, 74.00) &    (1.00, 92.00) &    (3.00, 115.00) &      (10.00, 186.00) \\
      (iv) \textbf{Sum of EN proj vector entries} &                 0.04 &             0.07 &             0.20 &             0.33 &             0.44 &             0.50 &             0.58 &             0.61 &              0.62 &                 0.60 \\
                                                  &               [0.26] &           [0.33] &           [0.45] &           [0.51] &           [0.51] &           [0.51] &           [0.58] &           [0.50] &            [0.38] &               [0.26] \\
                                                  &         (0.00, 0.93) &     (0.00, 1.00) &     (0.00, 1.00) &    (-0.08, 1.00) &    (-0.04, 1.00) &    (-0.20, 1.00) &    (-1.00, 1.00) &    (-0.81, 1.00) &     (-0.22, 1.00) &         (0.11, 0.91) \\
                     (v) \textbf{MktCap (in mil)} &             1,214.89 &           863.83 &           888.31 &         1,262.53 &         1,548.21 &         1,817.25 &         2,223.06 &         2,681.28 &          3,450.98 &             6,109.52 \\
                                                  &           [9,253.69] &       [7,449.02] &       [8,222.65] &      [11,497.41] &      [12,337.95] &      [13,904.73] &      [16,214.65] &      [16,726.44] &       [19,528.92] &          [26,175.47] \\
                                                  &     (5.55, 3,644.49) & (4.33, 2,283.37) & (3.99, 2,097.74) & (4.13, 2,828.01) & (4.59, 4,027.00) & (5.27, 5,272.39) & (6.47, 6,889.35) & (7.71, 9,005.63) & (9.48, 12,596.65) &   (14.86, 23,868.21) \\
                     (vi) \textbf{Book-to-market} &                 0.95 &             0.98 &             0.96 &             0.90 &             0.85 &             0.81 &             0.78 &             0.75 &              0.72 &                 0.68 \\
                                                  &               [1.17] &           [1.16] &           [1.17] &           [1.14] &           [1.07] &           [0.98] &           [0.94] &           [0.87] &            [0.84] &               [0.68] \\
                                                  &         (0.11, 2.45) &     (0.11, 2.57) &     (0.10, 2.54) &     (0.10, 2.38) &     (0.09, 2.24) &     (0.09, 2.12) &     (0.10, 2.01) &     (0.10, 1.89) &      (0.10, 1.79) &         (0.10, 1.64) \\
  (vii) \textbf{Dollar volume liquidity (in mil)} &                 6.96 &             4.67 &             5.46 &             8.71 &            11.39 &            13.70 &            16.80 &            20.39 &             26.66 &                44.11 \\
                                                  &              [56.09] &          [37.73] &          [50.05] &         [107.06] &         [107.76] &         [109.11] &         [114.71] &         [117.61] &          [127.33] &             [166.52] \\
                                                  &        (0.01, 23.11) &    (0.01, 13.79) &    (0.01, 14.42) &    (0.01, 24.60) &    (0.01, 37.15) &    (0.01, 50.20) &    (0.01, 64.98) &    (0.02, 86.54) &    (0.02, 123.23) &       (0.03, 198.25) \\
\bottomrule
\end{tabular}

		\caption{\textbf{Summary characteristics of elastic-net $\Rsquared$ sorted stocks}. %
			We project each stock's past twelve month's daily returns onto every other stock using the elastic-net estimator. Stocks are sorted into deciles based on their elastic-net $\Rsquared$ from the lowest (quantile 1, labelled ``Lo'') to highest (quantile 10, labelled ``Hi''). At the end of each month, we first compute the simple average of the stocks' characteristics within each bin. We then take the time-average of these averaged characteristics for each bin, and the displayed figure shows this cross-sectional and time-series average for each bin. Brackets show the standard deviation and the parentheses show the $(5^{\mathrm{th}}, 95^{\mathrm{th}})$ percentiles of the characteristics. The sampling period is from December 31, 1975 to November 30, 2020.
		}
		\label{tab:consolidated_characteristics_results}
\end{table}
\end{landscape}

\subsubsection{Time series correlations and replicates as hedging instruments} 
\label{sec:Correlations} 
How good are the replicates in mimicking the return behavior of the focals? While we will discuss at length the first moments of the focals and the replicates starting from Section~\ref{sec:SARP}, let's first investigate into their second cross-moments. For bins $k_1, k_2 \in \{ \mathrm{Lo}, 2, \ldots, \mathrm{Hi} \}$ and return type $m_1, m_2 \in \{\textrm{Foc}, \textrm{Rep} \}$ (recall \eqref{eq:PortfolioActual} and \eqref{eq:PortfolioGhost}), we track the time series $\bar{R}^{k_1}_{m_1, t}$ and the time series $\bar{R}^{k_2}_{m_2, t}$. The numbers in Table~\ref{tab:tbl_pairwise_ahead_ret_replicate_actual_corr} reports the sample correlation between the time series $\bar{R}^{k_1}_{m_1, t}$ and $\bar{R}^{k_2}_{m_2, t}$. The diagonal entries are most important: stocks with low (high) $\Rsquared$'s have low (high) correlations with their replicates. Specifically, the correlation between the returns of stocks with highest SAR (so they have the lowest $\Rsquared$) with their replicates is only $17.8\%$. In contrast, this correlation rises substantially along the diagonal of Table~\ref{tab:tbl_pairwise_ahead_ret_replicate_actual_corr} to $91.9\%$ for stocks with the lowest SAR (so highest $\Rsquared$). In all, for stocks with low SAR, the elastic-net replicate portfolio procedure (that is entirely based on past returns) do a fairly good job in matching the stock's one-month ahead return second moments. However, but expectedly, the replicates do a poor job in mimicking stocks with high SAR.

\begin{table}[htp]
	\centering
	\begin{tabular}{lcccccccccc}
\toprule
  &  Lo Rep &     2 &     3 &     4 &     5 &     6 &     7 &     8 &     9 &  Hi Rep \\
EN $\mathsf{R^2}$ &         &       &       &       &       &       &       &       &       &         \\
\midrule
Lo Foc            &   0.178 & 0.358 & 0.500 & 0.613 & 0.661 & 0.705 & 0.693 & 0.678 & 0.729 &   0.801 \\
2                 &   0.208 & 0.363 & 0.499 & 0.642 & 0.709 & 0.744 & 0.738 & 0.724 & 0.770 &   0.826 \\
3                 &   0.210 & 0.416 & 0.542 & 0.664 & 0.735 & 0.762 & 0.757 & 0.742 & 0.782 &   0.839 \\
4                 &   0.220 & 0.395 & 0.543 & 0.695 & 0.762 & 0.780 & 0.778 & 0.764 & 0.803 &   0.857 \\
5                 &   0.187 & 0.370 & 0.525 & 0.688 & 0.769 & 0.790 & 0.782 & 0.769 & 0.811 &   0.868 \\
6                 &   0.188 & 0.337 & 0.491 & 0.664 & 0.758 & 0.793 & 0.798 & 0.789 & 0.836 &   0.893 \\
7                 &   0.189 & 0.328 & 0.473 & 0.649 & 0.763 & 0.813 & 0.813 & 0.807 & 0.855 &   0.904 \\
8                 &   0.185 & 0.306 & 0.450 & 0.629 & 0.760 & 0.819 & 0.828 & 0.830 & 0.879 &   0.918 \\
9                 &   0.182 & 0.289 & 0.426 & 0.601 & 0.737 & 0.806 & 0.818 & 0.824 & 0.884 &   0.927 \\
Hi Foc            &   0.166 & 0.275 & 0.416 & 0.573 & 0.684 & 0.752 & 0.759 & 0.765 & 0.833 &   0.919 \\
\bottomrule
\end{tabular}

	\caption{ \textbf{Pairwise correlations of one-month ahead returns between stocks and their replicates. }%
	The $(k_1,k_2)$-th entry above represents the correlation of the time series one month ahead returns between the $k_1$-th elastic-net $\Rsquared$ sorted bin of the stocks and the $k_2$-th portfolio bin its replicate. The entries are expressed in decimals (e.g.\ 0.10 means 10\%). We form equal-weighted portfolios decile portfolios every month by projecting each stock's daily return over the past year onto every other stock using the elastic-net estimator. Stocks are sorted into deciles based on their elastic-net $\Rsquared$ from the lowest (quantile 1, labelled ``Lo'') to highest (quantile 10, labelled ``Hi''). The rows labelled ``Foc'' report the portfolio of the focal stocks \eqref{eq:Actual}. The columns labelled ``Rep'' report the replicate returns of the focal stock that are constructed out of the estimated elastic-net beta coefficients according to \eqref{eq:Ghost}. The sampling period is from January 31, 1976 to December 31, 2020. 
} 
	\label{tab:tbl_pairwise_ahead_ret_replicate_actual_corr} 
\end{table} 

\subsection{Main empirical result: Statistical arbitrage risk premium (SARP)}
\label{sec:SARP}
The univariate portfolio sorts of Table~\ref{tab:univariate_eql_wgt} show the main empirical result of this paper. Let's first discuss the returns of focal stocks in the first column of Table~\ref{tab:univariate_eql_wgt}. Stocks with the \emph{lowest} elastic-net $\Rsquared$ earn an excess return of 1.481\% ($t$-stat 5.298) per month, while stocks with the \emph{highest} elastic-net $\Rsquared$ earn an excess return of 0.771\% ($t$-stat 2.882). The \emph{difference} in returns between stocks with the lowest and highest elastic-net $\Rsquared$ is 0.710\% ($t$-stat 4.386). In other words, high SAR stocks have a substantially greater return than low SAR stocks.

\begin{table}[htp]
    \centering
		\begin{tabular}{lllllll}
\toprule
{} & \multicolumn{2}{l}{\textbf{Focal}} & \multicolumn{2}{l}{\textbf{Replicate}} & \multicolumn{2}{l}{\textbf{Foc - Rep}} \\
{} &           mean &      $t$ &               mean &       $t$ &               mean &      $t$ \\
EN $\mathsf{R^2}$ &                &          &                    &           &                    &          \\
\midrule
Lo                &          1.481 &  (5.298) &              0.115 &   (2.109) &              1.368 &  (4.929) \\
2                 &          1.167 &  (4.047) &              0.346 &   (2.767) &              0.818 &  (3.056) \\
3                 &          1.008 &  (3.408) &              0.383 &   (1.516) &              0.633 &  (2.459) \\
4                 &          0.941 &  (3.186) &              0.432 &   (1.451) &              0.511 &  (2.280) \\
5                 &          0.915 &  (3.196) &              0.360 &   (1.102) &              0.561 &  (2.634) \\
6                 &          0.889 &  (3.196) &              0.465 &   (1.400) &              0.431 &  (2.089) \\
7                 &          0.869 &  (3.177) &              0.724 &   (2.080) &              0.149 &  (0.720) \\
8                 &          0.891 &  (3.198) &              0.792 &   (2.303) &              0.093 &  (0.470) \\
9                 &          0.813 &  (2.943) &              0.655 &   (2.130) &              0.161 &  (1.112) \\
Hi                &          0.771 &  (2.882) &              0.503 &   (2.070) &              0.267 &  (2.581) \\
\textit{Lo - Hi}  &          0.710 &  (4.386) &             -0.387 &  (-1.662) &              1.101 &  (4.132) \\
\textit{Avg}      &          0.974 &  (3.584) &              0.478 &   (2.069) &              0.499 &  (3.391) \\
\bottomrule
\end{tabular}

		\caption{\textbf{\textcolor{red}{\uppercase{(Main Results)}} Univariate portfolio sort by elastic-net $\Rsquared$}. %
			We form equal-weighted decile portfolios every month by projecting each stock's past twelve month's daily returns onto every other stock using the elastic-net estimator. Stocks are sorted into deciles based on their elastic-net $\Rsquared$ from the lowest (quantile 1, labelled ``Lo'') to highest (quantile 10, labelled ``Hi''). The row labelled ``Lo - Hi'' is the monthly return difference between the ``Lo'' bin and the ``Hi'' bin. The row labelled ``Avg'' is the simple average of the monthly excess returns across the ten $k = \textrm{`Lo'}, 2, \ldots, \textrm{`Hi'}$ bins. The column labelled ``Focal'' reports the one-month ahead portfolio excess returns \eqref{eq:Actual}. The column labelled ``Replicate'' reports the one-month ahead excess returns of a portfolio that are constructed out of the estimated normalized elastic-net coefficients according to \eqref{eq:Ghost}. The column labelled ``Foc - Rep'' reports the one-month ahead returns of the portfolio of a long position in the focal stocks, and a short position in the corresponding replicates. We define the \emph{Statistical Arbitrage Risk Premium} (SARP) as the returns from ``Foc - Rep''. In addition, a stock with low EN $\Rsquared$ is said to have high \emph{Statistical Arbitrage Risk} (SAR), while a stock with high EN $\Rsquared$ is said to have low SAR. The ``mean'' column is reported in monthly percentage terms (e.g.\ 1.0 means 1\%). Robust \cite{Newey1987} $t$-statistics with six lags are reported in column ``$t$'' in parentheses. The sample period is monthly from January 31, 1976 to December 31, 2020. 
	}
    \label{tab:univariate_eql_wgt}
\end{table}

By conventional wisdom in the empirical asset pricing literature, these results already strongly hint that $\Rsquared$ is a potential priced factor in the cross-section. However, guided by our theoretical discussions of Section~\ref{sec:TheoreticalMotivation}, we are not just interested in testing the cross-sectional difference in the focal stocks (but see later in Section~\ref{sec:SARFactor} for a factor discussion). Rather, we are interested in empirically testing for the cross-sectional presence of SARP. Let's consider the empirical results of the replicates in the second column of Table~\ref{tab:univariate_eql_wgt}. The excess returns of the replicates are almost monotonically increasing, from $0.115\%$ ($t$-stat 2.109) per month in the \emph{lowest} $\Rsquared$ bin, to 0.503\% ($t$-stat 2.070) in the \emph{highest} $\Rsquared$ bin, even though their \emph{difference} -0.387\% ($t$-stat -1.662) is only weakly statistically significant. The excess returns of the replicates in some middle bins (e.g.\ bins 3 - 6) are statistically insignificant from zero. Other than these four middle bins, it appears the projection construction procedure does construct a replicate asset that has reasonable mean returns. 

Finally, we come to the main highlight of our paper: the third column of Table~\ref{tab:univariate_eql_wgt} proxies for the SARP. The focals minus replicates have an average return of 1.368\% ($t$-stat 4.929) per month for focal stocks with the \emph{lowest} $\Rsquared$'s, it is 0.267\% ($t$-stat 2.581) with the \emph{highest} $\Rsquared$'s, and the \emph{difference} of 1.101\% ($t$-stat 4.132) is highly statistically significant. This result is strong evidence of showing the key hypothesis of our theoretical discussions: assets with high SAR have high SARP, and assets with low SAR have low SARP. Beyond the aforementioned cross-sectional results, we also document the presence of an ``unconditional SARP''. The ``Avg'' portfolio in Table~\ref{tab:univariate_eql_wgt} takes the simple average of returns of all of the ten $k = \textrm{`Lo'}, 2, \ldots, \textrm{`Hi'}$ bins. We find that ``Avg'' enjoys a monthly SARP of 0.499\% ($t$-stat 3.391). 

In all, these empirical results show strong evidence in support of our core theoretical predictions: \emph{(i) an unconditional SARP exists and is positive for the average representative asset; and (ii) SARP is increasing with SAR: stocks with low (high) SAR earn a low (high) SARP in the cross-section.} 

\subsection{Controlling for other risk factors and characteristics} 
We present two sets of evidence to show our main result --- SARP is increasing in SAR --- from Table~\ref{tab:univariate_eql_wgt} is robust after controlling for risk factors and other characteristics.

\subsubsection{Risk adjusted returns on post-formation portfolios} 
\label{sec:FFRegOnPort} 
We show our results still persist even after adjusting for the Fama-French three factors (\cite{Fama1992cross, 1993_JFE_Fama}) and Fama-French five factors (\cite{Fama2015}). For each bin $k$ and each return $r_t^k  \in \{ \bar{R}^k_{\mathrm{Foc}, t}, \bar{R}^k_{\mathrm{Rep}, t}, \bar{R}^k_{\mathrm{LS}, t} \}$, we consider the following two time series factor model regressions: 
\begin{subequations}
	\begin{gather}
	r_{t}^k = \alpha^k + \beta^k_{\textrm{MktRF}} \textrm{MktRF}_t + \beta^k_{\textrm{SMB}}\textrm{SMB}_t + \beta^k_{\textrm{HML}}\textrm{HML}_t + \varepsilon_t^k, \label{eq:FF3} \\ 
	r_{t}^k = \alpha^k + \beta^k_{\textrm{MktRF}} \textrm{MktRF}_t + \beta^k_{\textrm{SMB}}\textrm{SMB}_t + \beta^k_{\textrm{HML}}\textrm{HML}_t + \beta^k_{\textrm{CMA}}\textrm{CMA}_t + \beta^k_{\textrm{RMW}} \textrm{RMW}_t + \varepsilon_t^k. 
	\label{eq:FF5}
	\end{gather} 
\end{subequations}
The regressors are well-known: ``MktRF'' is the market factor, ``SMB'' is the size factor, ``HML'' is the value factor, ``CMA'' is the investment factor, and ``RMW'' is the profitability factor. 

Table~\ref{tab:tbl_ff_on_port_consolidated_param_vals_eql_wgt} shows the FF3 and FF5 $\alpha$ estimates. Even controlling for known factors, the overall result that SARP is increasing in SAR remains robust. In particular, the FF3 $\alpha$ of SARP of stocks with the highest SAR is 0.573\% ($t$-stat 4.216) per month, while it is 0.020\% ($t$-stat 0.207) for stocks with the lowest SAR, and the cross-sectional difference is statistically significant at 0.553\% ($t$-stat 3.039). The cross-sectional $\alpha$ estimate of SARP using the FF5 model is analogous.

\begin{table}[htp]
		\centering 
		\begin{tabular}{lccccccccccc}
\toprule
{} & Lo EN $\mathsf{R^2}$ &        2 &         3 &         4 &         5 &         6 &         7 &         8 &         9 & Hi EN $\mathsf{R^2}$ & \textit{Lo - Hi} \\
\midrule
\textbf{Focal}          &                      &          &           &           &           &           &           &           &           &                      &                  \\
\phantom{0}FF3 $\alpha$ &                0.658 &    0.360 &     0.155 &     0.061 &     0.000 &    -0.063 &    -0.097 &    -0.100 &    -0.180 &               -0.219 &            0.877 \\
\phantom{0}             &              (5.338) &  (2.909) &   (1.166) &   (0.485) &   (0.001) &  (-0.722) &  (-1.500) &  (-1.581) &  (-2.437) &             (-2.376) &          (4.908) \\
\phantom{0}FF5 $\alpha$ &                0.700 &    0.383 &     0.140 &     0.044 &    -0.014 &    -0.051 &    -0.028 &     0.060 &     0.061 &               -0.033 &            0.733 \\
\phantom{0}             &              (5.684) &  (2.911) &   (1.012) &   (0.318) &  (-0.111) &  (-0.453) &  (-0.265) &   (0.529) &   (0.485) &             (-0.336) &          (4.217) \\
\textbf{Replicate}      &                      &          &           &           &           &           &           &           &           &                      &                  \\
\phantom{0}FF3 $\alpha$ &                0.087 &    0.198 &    -0.031 &    -0.200 &    -0.396 &    -0.353 &    -0.150 &    -0.061 &    -0.169 &               -0.239 &            0.326 \\
\phantom{0}             &              (1.526) &  (1.859) &  (-0.143) &  (-0.829) &  (-1.676) &  (-1.629) &  (-0.643) &  (-0.253) &  (-0.907) &             (-2.021) &          (2.962) \\
\phantom{0}FF5 $\alpha$ &                0.100 &    0.195 &    -0.052 &    -0.144 &    -0.209 &    -0.069 &     0.172 &     0.297 &     0.160 &               -0.062 &            0.162 \\
\phantom{0}             &              (1.631) &  (1.829) &  (-0.253) &  (-0.557) &  (-0.750) &  (-0.275) &   (0.643) &   (1.081) &   (0.758) &             (-0.507) &          (1.399) \\
\textbf{Foc - Rep}      &                      &          &           &           &           &           &           &           &           &                      &                  \\
\phantom{0}FF3 $\alpha$ &                0.573 &    0.163 &     0.197 &     0.264 &     0.403 &     0.299 &     0.060 &    -0.045 &    -0.008 &                0.020 &            0.553 \\
\phantom{0}             &              (4.216) &  (1.142) &   (1.040) &   (1.271) &   (1.913) &   (1.437) &   (0.275) &  (-0.209) &  (-0.055) &              (0.207) &          (3.039) \\
\phantom{0}FF5 $\alpha$ &                0.602 &    0.187 &     0.199 &     0.193 &     0.203 &     0.031 &    -0.193 &    -0.242 &    -0.098 &                0.028 &            0.574 \\
\phantom{0}             &              (4.330) &  (1.231) &   (1.041) &   (0.900) &   (0.909) &   (0.148) &  (-0.854) &  (-1.081) &  (-0.618) &              (0.281) &          (3.149) \\
\bottomrule
\end{tabular}

		\caption{\textbf{Fama-French 3 and 5 factor regressions on portfolios sorted by elastic-net $\Rsquared$.} %
			We run the Fama-French three \eqref{eq:FF3} and five \eqref{eq:FF5} factor time series regressions onto the elastic-net $\Rsquared$ sorted decile portfolios. Stocks are sorted into deciles based on their elastic-net $\Rsquared$ from the lowest (quantile 1, labelled ``Lo'') to highest (quantile 10, labelled ``Hi''). The column labelled ``Lo - Hi'' is the monthly return difference between the ``Lo'' bin and the ``Hi'' bin. The row labelled ``Avg'' is the simple average of the monthly excess returns across the ten $k = \textrm{`Lo'}, 2, \ldots, \textrm{`Hi'}$ bins. The group labelled ``Focal'' reports the one-month ahead portfolio excess returns \eqref{eq:Actual}. The group labelled ``Replicate'' reports the one-month ahead excess returns of a portfolio that are constructed out of the estimated normalized elastic-net beta coefficients according to \eqref{eq:Ghost}. The group labelled ``Foc - Rep'' reports the one-month ahead returns of the portfolio of a long position in the focal stocks, and a short position in the corresponding replicates. We define the \emph{Statistical Arbitrage Risk Premium} (SARP) as the returns from ``Foc - Rep''. In addition, a stock with low EN $\Rsquared$ is said to have high \emph{Statistical Arbitrage Risk} (SAR), while a stock with high EN $\Rsquared$ is said to have low SAR. The estimated values are reported in monthly percentage points (e.g.\ 1.0 means 1\%) and we report the \cite{Newey1987} $t$-statistics with six lags in parentheses. The sample period is from January 31, 1976 to December 31, 2020.
		}
		\label{tab:tbl_ff_on_port_consolidated_param_vals_eql_wgt}
\end{table} 

\subsection{Bivariate dependent sorts}
\label{sec:BivariateSorts}
We show our main result remains robust after we control for various stock characteristics. At the end of each month, stocks are first sorted by a characteristic into quintile portfolios. Then for a given characteristic and within each of its five portfolios, we further sort stocks based on the elastic-net $\Rsquared$ into deciles bins. Each of these ten elastic-net $\Rsquared$ bins are then averaged over their respective five characteristic portfolios. In all, the ten resulting elastic-net $\Rsquared$ bins represent returns that control for a particular characteristic. All portfolios are equal-weighted. 

Let's describe the characteristics. \emph{MktCap} is market capitalization and \emph{B/M} is book-to-market, and their quintiles are calculated using the NYSE breakpoints. \emph{IdioVol} is \cite{2006_JF_Ang}'s idiosyncratic volatility; the idiosyncratic volatility of stock $i$ at month $t-1$ is the standard deviation of the residuals, arising from the OLS regression of past one year's daily returns leading up until month $t - 1$ onto the Fama-French three factors. \emph{TotalVol} is total volatility; it is the sample standard deviation of a stock $i$'s past twelve months of daily returns up until month $t - 1$. \emph{IdioSkew} is idiosyncratic skewness; it is calculated as the sample Pearson's moment of coefficient skewness of the residuals of the OLS regression of a stock's past twelve months' daily returns up until monh $t - 1$ onto the Fama-French three factors (with intercept). \emph{TotalSkew} is total skewness; it is calculated as the sample Pearson's moment of coefficient skewness of a stock's past twelve months' daily returns up until month $t - 1$. \emph{AmihudIlliq} is \cite{Amihud2002}'s illiquidity measure; it is defined as $\textrm{AmihudIlliq}_{i,t-1} := \frac{1}{D_{i,t-1}} \sum_{d=1}^{D_{i,t-1}} \frac{\abs{R_{i,d}}}{\textrm{VOLD}_{i,d}}$, where $D_{i,t-1}$ is the total number of trading days of stock $i$ in the past twelve months leading up to month $t - 1$ and \textrm{VOLD} was as defined in Section~\ref{sec:CharacteristicsStatistics}. \emph{Mom} is momentum; we define momentum of stock $i$ at the end of month $t - 1$ (i.e.\ the end of the estimation period) as the return of the stock during the 11-month period covering months $t - 12$ through $t - 2$. \emph{STR} is \cite{Jegadeesh1990} and \cite{Lehmann1990}'s short-term reversal; it is defined as $\textrm{STR}_{i,t-1} = 100 \times R_{i,t-1}$. 

Table~\ref{tab:consolidated_bivariate_eql_wgt} shows the results. We omit showing the results on the replicates for brevity. Overall, these results that control for the aforementioned characteristics are consistent with the unconditional results of Table~\ref{tab:univariate_eql_wgt}. We pay special attention our results after controlling for size, value, VOLD, idiosyncratic volatility and total volatility. In particular, even though Table~\ref{tab:consolidated_characteristics_results} shows a size, value and VOLD tilt for the elastic-net $\Rsquared$ sorted stocks, the results here show that SARP is still persistent after we control for these characteristics. 

\begin{table}[htp]
    \begin{tabular}{llccccccccccc}
\toprule
         \phantom{0} &        \phantom{1} & Lo EN $\mathsf{R^2}$ &       2 &       3 &       4 &       5 &       6 &       7 &       8 &       9 & Hi EN $\mathsf{R^2}$ & \textit{Lo - Hi} \\
\midrule
     \textbf{MktCap} &     \textbf{Focal} &                1.389 &   1.101 &   0.995 &   0.937 &   0.811 &   0.826 &   0.880 &   0.954 &   0.911 &                0.921 &            0.468 \\
                     &                    &              (5.527) & (4.335) & (3.829) & (3.521) & (3.018) & (2.974) & (3.078) & (3.162) & (2.997) &              (2.917) &          (3.070) \\
                     & \textbf{Foc - Rep} &                1.085 &   0.680 &   0.616 &   0.642 &   0.420 &   0.481 &   0.236 &   0.180 &   0.205 &                0.404 &            0.681 \\
                     &                    &              (4.863) & (3.207) & (3.380) & (3.691) & (2.255) & (2.726) & (1.168) & (0.989) & (1.438) &              (3.600) &          (3.737) \\
                     &                    &                      &         &         &         &         &         &         &         &         &                      &                  \\
        \textbf{B/M} &     \textbf{Focal} &                1.434 &   1.150 &   0.954 &   1.002 &   0.931 &   0.920 &   0.911 &   0.936 &   0.926 &                0.882 &            0.552 \\
                     &                    &              (5.074) & (4.074) & (3.250) & (3.586) & (3.346) & (3.324) & (3.347) & (3.422) & (3.461) &              (3.404) &          (3.835) \\
                     & \textbf{Foc - Rep} &                1.293 &   0.820 &   0.577 &   0.608 &   0.524 &   0.461 &   0.198 &   0.135 &   0.242 &                0.353 &            0.941 \\
                     &                    &              (4.716) & (3.268) & (2.446) & (3.009) & (2.556) & (2.405) & (0.972) & (0.660) & (1.524) &              (3.395) &          (3.516) \\
                     &                    &                      &         &         &         &         &         &         &         &         &                      &                  \\
    \textbf{IdioVol} &     \textbf{Focal} &                1.416 &   1.090 &   1.015 &   0.952 &   0.985 &   0.855 &   0.945 &   0.923 &   0.827 &                0.781 &            0.635 \\
                     &                    &              (5.453) & (4.285) & (3.851) & (3.549) & (3.651) & (3.040) & (3.234) & (3.116) & (2.749) &              (2.451) &          (4.031) \\
                     & \textbf{Foc - Rep} &                1.196 &   0.641 &   0.665 &   0.610 &   0.663 &   0.402 &   0.241 &   0.159 &   0.159 &                0.300 &            0.896 \\
                     &                    &              (4.830) & (2.659) & (3.214) & (2.994) & (3.401) & (2.163) & (1.205) & (0.836) & (1.099) &              (2.618) &          (4.371) \\
                     &                    &                      &         &         &         &         &         &         &         &         &                      &                  \\
   \textbf{TotalVol} &     \textbf{Focal} &                1.422 &   1.124 &   1.004 &   0.977 &   0.960 &   0.859 &   0.921 &   0.893 &   0.849 &                0.777 &            0.645 \\
                     &                    &              (5.244) & (4.252) & (3.642) & (3.584) & (3.463) & (3.043) & (3.216) & (3.091) & (2.921) &              (2.657) &          (4.557) \\
                     & \textbf{Foc - Rep} &                1.276 &   0.676 &   0.639 &   0.633 &   0.604 &   0.436 &   0.160 &   0.141 &   0.174 &                0.295 &            0.981 \\
                     &                    &              (4.888) & (2.772) & (2.870) & (3.148) & (3.103) & (2.379) & (0.788) & (0.743) & (1.194) &              (2.851) &          (4.301) \\
                     &                    &                      &         &         &         &         &         &         &         &         &                      &                  \\
\textbf{AmihudIlliq} &     \textbf{Focal} &                1.367 &   1.103 &   0.973 &   0.992 &   0.866 &   0.850 &   0.950 &   0.871 &   0.979 &                0.932 &            0.434 \\
                     &                    &              (5.436) & (4.261) & (3.747) & (3.653) & (3.175) & (3.116) & (3.362) & (2.970) & (3.390) &              (3.258) &          (3.071) \\
                     & \textbf{Foc - Rep} &                0.996 &   0.730 &   0.583 &   0.724 &   0.510 &   0.473 &   0.300 &   0.099 &   0.295 &                0.419 &            0.577 \\
                     &                    &              (4.546) & (3.655) & (3.142) & (4.076) & (2.826) & (2.697) & (1.478) & (0.508) & (2.042) &              (3.906) &          (3.197) \\
                     &                    &                      &         &         &         &         &         &         &         &         &                      &                  \\
        \textbf{Mom} &     \textbf{Focal} &                1.428 &   1.114 &   1.103 &   0.976 &   0.975 &   0.882 &   0.894 &   0.895 &   0.775 &                0.703 &            0.725 \\
                     &                    &              (5.151) & (3.901) & (3.856) & (3.429) & (3.470) & (3.167) & (3.267) & (3.288) & (2.779) &              (2.626) &          (5.282) \\
                     & \textbf{Foc - Rep} &                1.311 &   0.739 &   0.725 &   0.561 &   0.611 &   0.432 &   0.155 &   0.110 &   0.092 &                0.216 &            1.095 \\
                     &                    &              (4.803) & (2.841) & (3.048) & (2.611) & (2.914) & (2.301) & (0.763) & (0.537) & (0.604) &              (2.168) &          (4.309) \\
                     &                    &                      &         &         &         &         &         &         &         &         &                      &                  \\
        \textbf{STR} &     \textbf{Focal} &                1.461 &   1.171 &   0.983 &   0.945 &   0.881 &   0.833 &   0.888 &   0.901 &   0.834 &                0.822 &            0.638 \\
                     &                    &              (5.252) & (4.069) & (3.464) & (3.199) & (3.146) & (3.038) & (3.187) & (3.314) & (3.010) &              (3.027) &          (4.213) \\
                     & \textbf{Foc - Rep} &                1.332 &   0.825 &   0.616 &   0.528 &   0.526 &   0.334 &   0.162 &   0.095 &   0.195 &                0.320 &            1.012 \\
                     &                    &              (4.893) & (3.235) & (2.525) & (2.408) & (2.631) & (1.724) & (0.792) & (0.469) & (1.377) &              (3.160) &          (3.972) \\
                     &                    &                      &         &         &         &         &         &         &         &         &                      &                  \\
       \textbf{VOLD} &     \textbf{Focal} &                1.341 &   1.103 &   0.985 &   0.936 &   0.871 &   0.831 &   0.951 &   0.929 &   0.973 &                0.949 &            0.392 \\
                     &                    &              (5.153) & (4.142) & (3.700) & (3.427) & (3.153) & (3.020) & (3.337) & (3.246) & (3.491) &              (3.511) &          (2.782) \\
                     & \textbf{Foc - Rep} &                1.033 &   0.745 &   0.610 &   0.603 &   0.528 &   0.459 &   0.247 &   0.170 &   0.303 &                0.423 &            0.611 \\
                     &                    &              (4.711) & (3.686) & (3.261) & (3.487) & (3.006) & (2.673) & (1.234) & (0.842) & (2.017) &              (4.029) &          (3.211) \\
                     &                    &                      &         &         &         &         &         &         &         &         &                      &                  \\
   \textbf{IdioSkew} &     \textbf{Focal} &                1.451 &   1.148 &   1.041 &   0.943 &   0.903 &   0.856 &   0.843 &   0.898 &   0.864 &                0.787 &            0.664 \\
                     &                    &              (5.408) & (4.159) & (3.494) & (3.180) & (3.218) & (3.098) & (2.983) & (3.195) & (3.035) &              (2.885) &          (4.310) \\
                     & \textbf{Foc - Rep} &                1.300 &   0.773 &   0.711 &   0.538 &   0.563 &   0.375 &   0.202 &   0.061 &   0.169 &                0.287 &            1.013 \\
                     &                    &              (4.938) & (2.915) & (3.234) & (2.558) & (2.986) & (1.928) & (1.057) & (0.293) & (1.172) &              (2.773) &          (4.095) \\
                     &                    &                      &         &         &         &         &         &         &         &         &                      &                  \\
  \textbf{TotalSkew} &     \textbf{Focal} &                1.437 &   1.156 &   1.032 &   0.982 &   0.950 &   0.804 &   0.842 &   0.908 &   0.831 &                0.794 &            0.643 \\
                     &                    &              (5.438) & (4.175) & (3.478) & (3.397) & (3.348) & (2.926) & (2.973) & (3.207) & (2.898) &              (2.897) &          (4.267) \\
                     & \textbf{Foc - Rep} &                1.271 &   0.758 &   0.685 &   0.613 &   0.596 &   0.342 &   0.175 &   0.064 &   0.185 &                0.290 &            0.980 \\
                     &                    &              (4.982) & (2.894) & (3.107) & (3.029) & (3.214) & (1.774) & (0.851) & (0.325) & (1.296) &              (2.908) &          (4.118) \\
                     &                    &                      &         &         &         &         &         &         &         &         &                      &                  \\
\bottomrule
\end{tabular}

		\caption{\textbf{Dependent bivariate sorts by characteristics and elastic-net $\Rsquared$. }%
			This table reports mean returns in monthly percentage points (e.g.\ 1.0 means 1\%) and \cite{Newey1987} robust $t$-statistics with 6 lags in parentheses. We perform a bivariate dependent sort. Each month, we first sort stocks based on a characteristic (i.e.\ size, book-to-market, idiosyncratic volatility, total volatility, Amihud's illiquidity, momentum, short-term reversal, dollar volume liquidity, idiosyncratic skewness, and total skewness; see Section~\ref{sec:BivariateSorts} for details and references on these characteristics) into quintiles portfolios. For a given characteristic and within each of its five portfolios, we sort stocks based on elastic-net $\Rsquared$ into decile bins, from the lowest (quantile 1, labelled ``Lo'') to highest (quantile 10, labelled ``Hi''). The column labelled ``Lo - Hi'' is the monthly return difference between the ``Lo'' bin and the ``Hi'' bin. These ten elastic-net $\Rsquared$ bins are then averaged over each of the five characteristic portfolios. Thus these ten elastic-net $\Rsquared$ bins represent returns that control for a particular characteristic. All portfolios are equal-weighted. The sample period is monthly from January 31, 1976 to December 31, 2020.
		}
		\label{tab:consolidated_bivariate_eql_wgt}
\end{table} 

\subsection{Statistical Arbitrage Risk factor (SAR factor)}
\label{sec:SARFactor}
The main empirical objective of this paper is to show assets' SARP increases with SAR. However as hinted in the main results Section~\ref{sec:SARP}, conventional empirical asset pricing results would suggest that SAR is a priced factor. We can show the following corollary factor result. Let us define the \emph{Statistical Arbitrage Risk factor (SAR factor)} with returns at month $t$ as
\begin{equation}
	R_{\mathrm{SAR},t}
	:= \bar{R}^{\mathrm{Lo}}_{\mathrm{LS}, t}  - \bar{R}^{\mathrm{Hi}}_{\mathrm{LS}, t}.
	\label{eq:SARFactor}
\end{equation}
where recall the definition of $\bar{R}^k_{\mathrm{LS}, t}$ in \eqref{eq:PortfolioLS}. In other words, the SAR factor is simply the difference in the SARP of the high SAR and low SAR stocks.

Let's investigate the price of risk $\lambda_{\mathrm{SAR}}$ of our SAR factor by the classical \cite{fama1973risk} regressions. We run the \cite{fama1973risk} with four different models: (i) SAR factor + CAPM; (ii) SAR factor + FF3 + momentum; (iii) SAR factor + FF5 + momentum; and (iv) FF5 + momentum as a benchmark. The data of the momentum factor is available from Kenneth French's website. Table~\ref{tab:fmreg_sel_ahead_ret_actual_minus_replicate_eql_wgt}(a) shows the correlations of our SAR factor along with the other asset pricing factors. Next is the choice of test assets. We first evaluate the price of risk of our SAR factor on the classical 25 portfolios formed on size and book-to-market ($5 \times 5)$. In addition, given our projection and replicate construction procedure of Section~\ref{sec:Methodology} is explicitly dependent on past returns, it is prudent to evaluate our SAR factor on test assets that are also sorted along a past return dependent characteristic. To this end, we will also consider the 25 portfolios formed on size and momentum ($5 \times 5$) and the 25 portfolios formed on size and residual variance ($5 \times 5)$. Finally, we also evaluate our SAR factor on portfolios that are further sorted along corporate fundamentals, and consider the 25 portfolios formed on book-to-market and investments ($5 \times 5$), and the 36 portfolios formed on size, operating profitability and investments ($2 \times 4 \times 4$). All test portfolios are equal-weighted and are obtained from French's website. Table~\ref{tab:fmreg_sel_ahead_ret_actual_minus_replicate_eql_wgt} shows the main result of this section and the focus is the value of $\lambda_{\mathrm{SAR}}$. The SAR factor has a positive price of risk, is statistically significant, and is robust across the various model specifications and test assets. Focusing on the SAR factor + FF5 + momentum model, we see the price of risk estimate for the SAR factor range from 0.990\% to 1.631\% per month, depending on the test asset. 

\begin{table}[htp]
	\scriptsize
		\centering 
			\begin{tabular}{lccccccccc}
\toprule
                                                                   Test asset & $\lambda_{\mathrm{SAR}}$ & $\lambda_{\mathrm{MktRF}}$ & $\lambda_{\mathrm{SMB}}$ & $\lambda_{\mathrm{HML}}$ & $\lambda_{\mathrm{RMW}}$ & $\lambda_{\mathrm{CMA}}$ & $\lambda_{\mathrm{MOM}}$ &    $\chi^2$ \\
\midrule
                                         \textbf{Size and B/M ($5 \times 5$)} &                          &                            &                          &                          &                          &                          &                          &             \\
                                                                              &                          &                      0.993 &                    0.445 &                    0.215 &                   -0.171 &                    0.326 &                    2.286 &       96.91 \\
                                                                              &                          &                    (4.695) &                  (3.308) &                  (1.575) &                 (-0.724) &                  (1.861) &                  (4.122) & [p = 0.000] \\
                                                                              &                    0.652 &                      0.777 &                          &                          &                          &                          &                          &       64.66 \\
                                                                              &                  (2.354) &                    (3.817) &                          &                          &                          &                          &                          & [p = 0.000] \\
                                                                              &                    1.099 &                      0.997 &                    0.438 &                    0.215 &                          &                          &                    2.489 &       56.24 \\
                                                                              &                  (3.989) &                    (4.774) &                  (3.267) &                  (1.589) &                          &                          &                  (4.449) & [p = 0.000] \\
                                                                              &                    1.224 &                      0.861 &                    0.434 &                    0.184 &                    0.366 &                   -0.037 &                    1.876 &      161.15 \\
                                                                              &                  (4.288) &                    (4.120) &                  (3.234) &                  (1.348) &                  (1.821) &                 (-0.196) &                  (3.577) & [p = 0.000] \\
                                    \textbf{Size and Momentum ($5 \times 5$)} &                          &                            &                          &                          &                          &                          &                          &             \\
                                                                              &                          &                      0.717 &                    0.308 &                    1.053 &                   -0.467 &                    0.894 &                    0.615 &       64.50 \\
                                                                              &                          &                    (3.618) &                  (2.203) &                  (3.397) &                 (-1.763) &                  (2.907) &                  (3.209) & [p = 0.000] \\
                                                                              &                    0.948 &                      0.666 &                          &                          &                          &                          &                          &       88.03 \\
                                                                              &                  (3.492) &                    (3.291) &                          &                          &                          &                          &                          & [p = 0.000] \\
                                                                              &                    1.097 &                      0.674 &                    0.334 &                    0.782 &                          &                          &                    0.616 &       59.11 \\
                                                                              &                  (4.010) &                    (3.403) &                  (2.486) &                  (3.707) &                          &                          &                  (3.215) & [p = 0.000] \\
                                                                              &                    1.031 &                      0.688 &                    0.306 &                    0.972 &                   -0.206 &                    0.461 &                    0.618 &      229.47 \\
                                                                              &                  (3.633) &                    (3.462) &                  (2.192) &                  (3.121) &                 (-0.738) &                  (1.370) &                  (3.223) & [p = 0.000] \\
                           \textbf{Size and Residual Variance ($5 \times 5$)} &                          &                            &                          &                          &                          &                          &                          &             \\
                                                                              &                          &                      0.555 &                    0.257 &                    1.216 &                    0.052 &                    0.295 &                   -0.541 &       39.07 \\
                                                                              &                          &                    (2.809) &                  (1.705) &                  (5.292) &                  (0.340) &                  (0.968) &                 (-0.822) & [p = 0.027] \\
                                                                              &                    0.406 &                      0.832 &                          &                          &                          &                          &                          &      581.88 \\
                                                                              &                  (1.462) &                    (4.134) &                          &                          &                          &                          &                          & [p = 0.000] \\
                                                                              &                    1.898 &                      0.710 &                    0.618 &                    0.775 &                          &                          &                    3.380 &       84.35 \\
                                                                              &                  (4.998) &                    (3.595) &                  (3.967) &                  (3.601) &                          &                          &                  (4.358) & [p = 0.000] \\
                                                                              &                    1.631 &                      0.678 &                    0.434 &                    1.037 &                    0.225 &                   -0.453 &                    2.379 &       74.61 \\
                                                                              &                  (4.507) &                    (3.450) &                  (2.835) &                  (4.500) &                  (1.443) &                 (-1.348) &                  (3.462) & [p = 0.000] \\
                                   \textbf{B/M and Investment ($5 \times 5$)} &                          &                            &                          &                          &                          &                          &                          &             \\
                                                                              &                          &                      0.921 &                    0.177 &                    0.638 &                   -0.202 &                    0.826 &                    0.751 &       11.53 \\
                                                                              &                          &                    (3.121) &                  (0.533) &                  (3.372) &                 (-0.875) &                  (7.591) &                  (1.496) & [p = 0.985] \\
                                                                              &                    1.021 &                      0.681 &                          &                          &                          &                          &                          &       72.78 \\
                                                                              &                  (3.666) &                    (3.240) &                          &                          &                          &                          &                          & [p = 0.000] \\
                                                                              &                    1.868 &                      1.190 &                   -0.083 &                    0.490 &                          &                          &                    2.615 &       84.89 \\
                                                                              &                  (5.428) &                    (4.846) &                 (-0.328) &                  (3.084) &                          &                          &                  (4.581) & [p = 0.000] \\
                                                                              &                    1.000 &                      0.960 &                    0.106 &                    0.607 &                   -0.116 &                    0.807 &                    0.892 &       18.96 \\
                                                                              &                  (2.573) &                    (3.262) &                  (0.326) &                  (3.088) &                 (-0.458) &                  (7.129) &                  (1.567) & [p = 0.754] \\
\textbf{Size, Operating profitability and Investment ($2 \times 4 \times 4$)} &                          &                            &                          &                          &                          &                          &                          &             \\
                                                                              &                          &                      0.820 &                    0.425 &                    0.534 &                   -0.007 &                    0.525 &                    1.317 &      159.07 \\
                                                                              &                          &                    (3.998) &                  (2.726) &                  (2.677) &                 (-0.061) &                  (5.034) &                  (3.020) & [p = 0.000] \\
                                                                              &                    0.766 &                      0.836 &                          &                          &                          &                          &                          &      223.96 \\
                                                                              &                  (2.772) &                    (4.143) &                          &                          &                          &                          &                          & [p = 0.000] \\
                                                                              &                    1.169 &                      0.786 &                    0.236 &                    0.879 &                          &                          &                    1.106 &      491.33 \\
                                                                              &                  (4.145) &                    (3.801) &                  (1.556) &                  (5.312) &                          &                          &                  (2.287) & [p = 0.000] \\
                                                                              &                    0.990 &                      0.808 &                    0.339 &                    0.668 &                    0.019 &                    0.477 &                    1.232 &      136.89 \\
                                                                              &                  (3.376) &                    (3.943) &                  (2.077) &                  (3.093) &                  (0.166) &                  (4.445) &                  (2.814) & [p = 0.000] \\
\bottomrule
\end{tabular}

		\caption{\textbf{SAR factor price of risk}. %
			This table shows the market price of risk $\lambda_{\mathrm{SAR}}$ in monthly percentage points (i.e. 1.0 means 1\%) of the SAR factor and various other factors, and where the parentheses show the \cite{fama1973risk} $t$-statistics. The last column shows the $\chi^2$ statistic and the brackets show its $p$-value. We define our \emph{Statistical Arbitrage Risk factor} (SAR factor) in \eqref{eq:SARFactor}; the factor return $R_{\mathrm{SAR},t}$ is a result of a long position on the Lo elastic-net $\Rsquared$ bin with returns $\bar{R}^{\mathrm{Lo}}_{\mathrm{LS}, t}$ and a short position on the Hi elastic $\Rsquared$ bin with returns $\bar{R}^{\mathrm{Hi}}_{\mathrm{LS}, t}$. We focus on five equal-weighted tests assets: the 25 portfolios formed by the 5 size and 5 book-to-market bins; the 25 portfolios formed by the 5 size and 5 momentum bins; the 25 portfolios formed by size and \cite{2006_JF_Ang} residual variance bins; 25 portfolios formed by 5 book-to-market and 5 investment bins; and 36 portfolios formed by 2 size, 4 operating profitability and 4 investment bins. We use the \cite{fama1973risk} two-pass procedure by first running a monthly time series regression of the test assets onto the proposed factor models to obtain the betas, and then run a cross-sectional regression to obtain the prices of risk. The sampling period is monthly from January 31, 1976 to December 31, 2020. 
		}
		\label{tab:fmreg_sel_ahead_ret_actual_minus_replicate_eql_wgt}
\end{table} 

\subsubsection{Investment strategy}
\label{sec:InvestmentStrategy}
\begin{figure}
	\centering
	\includegraphics[scale = 0.55]{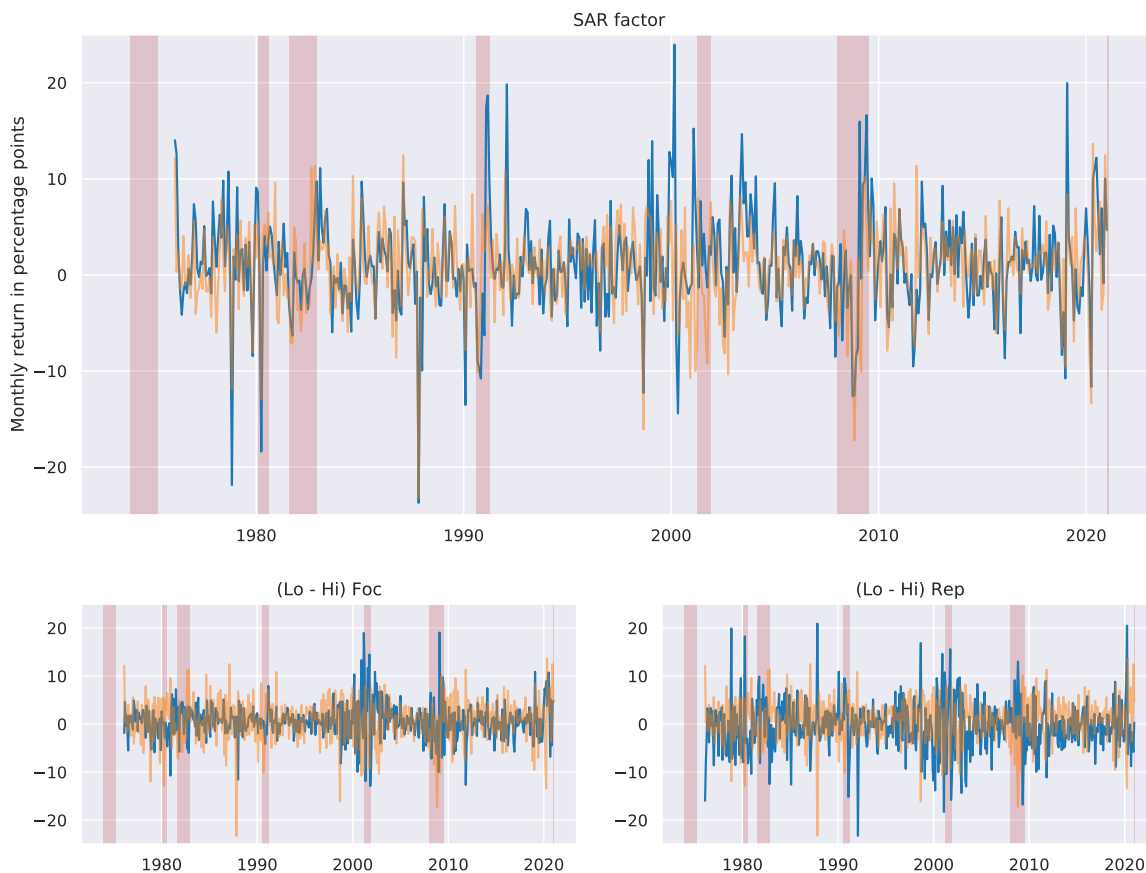}
	\caption{\textbf{Time series plot of monthly returns of the SAR factor}. %
		The top panel plots the monthly returns of the SAR factor of \eqref{eq:SARFactor} in blue, while the orange line is that of the market factor for comparison. Returns are expressed in monthly percentage points (e.g. 1.0 means 1\%). The bottom left panel plots the strategy ``(Lo - Hi) Foc'', while the bottom right panel plots the strategy ``(Lo - Hi) Rep''. The red shaded regions are NBER recession dates. The sampling period is monthly from January 31, 1976 to December 31, 2020.
	}
	\label{fig:time_series_plot_eql_wgt}
\end{figure}

\begin{figure}
	\centering
	\includegraphics[scale = 0.58]{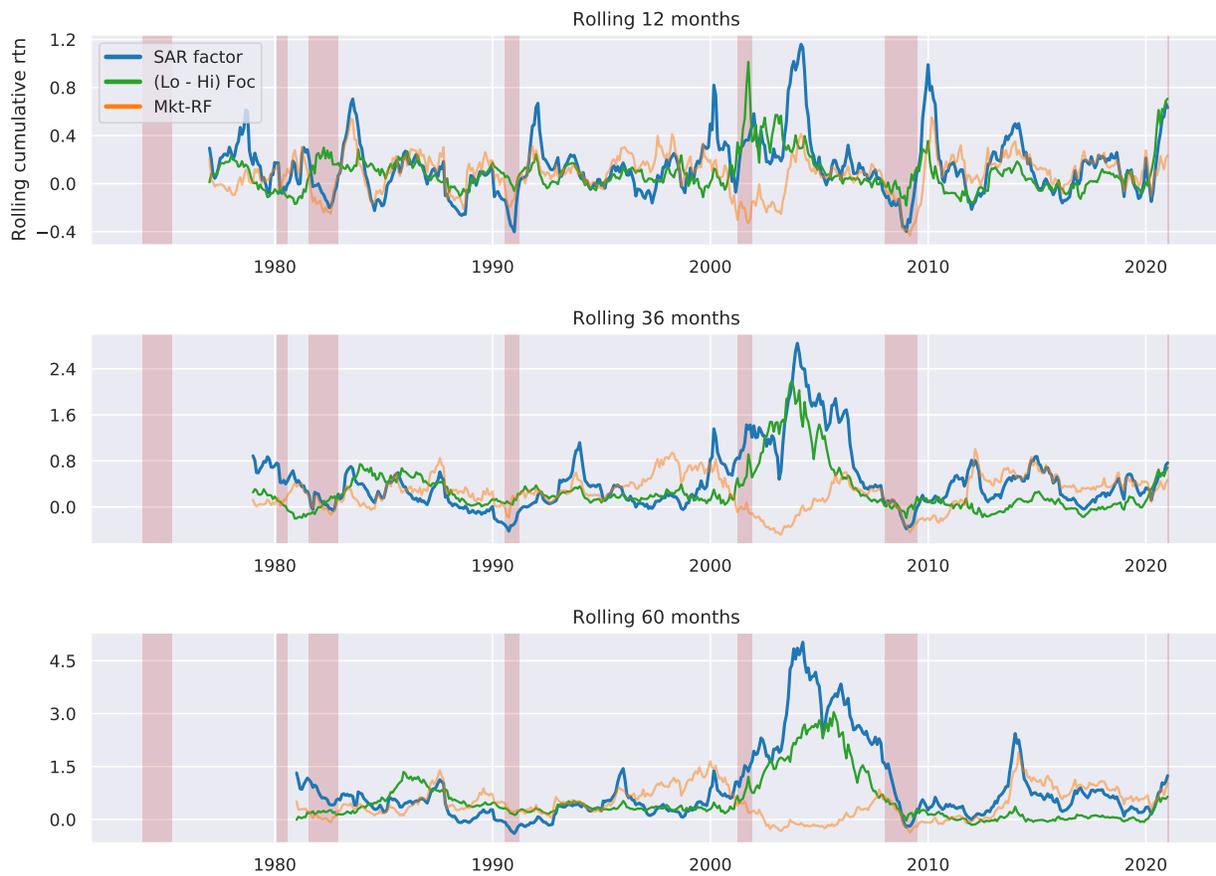}
	\caption{\textbf{Rolling window cumulative returns of the SAR factor}. %
		We plot the cumulative returns of rolling $n =$ 12, 36 and 60 months of an initial $\$1$ investment of the SAR factor, the ``(Lo - Hi) Foc'' strategy, and the market factor for comparison (e.g.\ 1.5 means a total of return of 150\% over the past $n$ many months). The red shaded regions are NBER recession dates. The sampling period is monthly from January 31, 1976 to December 31, 2020.
	}
	\label{fig:rollingwindow_rtn_plot_eql_wgt}
\end{figure}

The SAR factor is clearly a tradable portfolio. Figure~\ref{tab:CumRtn}(a) shows the cumulative returns (not inflation adjusted) from December 31, 1975 to December 31, 2020 of an initial $\$100$ investment on our SAR factor and other factors, and Figure~\ref{tab:CumRtn}(b) plots the log cumulative returns. The strategy ``(Lo - Hi) Foc'' is the cumulative returns from $\bar{R}^{\mathrm{Lo}}_{\mathrm{Foc}, t} - \bar{R}^{\mathrm{Hi}}_{\mathrm{Foc}, t}$; the strategy ``(Lo - Hi) Rep'' is that of $\bar{R}^{\mathrm{Lo}}_{\mathrm{Rep}, t} - \bar{R}^{\mathrm{Hi}}_{\mathrm{Rep}, t}$; and ``SAR factor'' is that of \eqref{eq:SARFactor}. Following the ``(Lo - Hi) Foc'' strategy will result in a $\$3,177.63$ cumulative return on December 31, 2020, ``(Lo - Hi) Rep'' will result in $\$5.70$ and SAR factor will result in $\$18,219.66$. The amazing performance of SAR factor is firstly because the ``(Lo - Hi) Foc'' strategy itself already has a high return. More importantly, however, the SAR factor strategy simultaneously takes advantage of both the low returns of ``(Lo - Hi) Rep'' as leverage and also its high correlations with the focal stocks. As a matter of comparison, investing into the risk-free asset would result in the cumulative return of $\$687.29$; the market factor returns $\$2,505.58$; the SMB factor $\$264.59$; the HML factor $\$220.58$; the CMA factor $\$318.83$; the RMW factor $\$467.17$; and the MOM factor $\$1,547.36$. See Figure~\ref{fig:time_series_plot_eql_wgt} for a plot of the monthly returns of the SAR factor. See Figure~\ref{fig:rollingwindow_rtn_plot_eql_wgt} for a plot of the rolling cumulative returns of the SAR factor.

\begin{table}[htp]
	\centering
	\begin{minipage}[h]{\textwidth}
		\centering
		\begin{subfigure}{1\textwidth}
			\begin{tabular}{lrcccccccc}
\toprule
\phantom{Asset type} & Cum.\ return &  Ann. Sharpe ratio &   Mean &   Std & Skewness & Kurtosis &  (5, 95)-th pct & $\textrm{ARMA}(1, \cdot)$ coef & $\textrm{ARMA}(\cdot, 1)$ coef \\
\midrule
       (Lo - Hi) Foc &     3,177.63 &              0.329 &   0.71 & 3.699 &    0.315 &    3.523 & (-5.325, 6.239) &                 -0.95 (-11.87) &                   0.93 (10.53) \\
       (Lo - Hi) Rep &         5.70 &             -0.489 & -0.387 & 5.308 &    0.053 &    2.258 & (-8.898, 7.904) &                  -0.09 (-0.29) &                    0.20 (0.68) \\
          SAR factor &    18,219.66 &              0.495 &  1.101 & 5.169 &     0.07 &    3.456 & (-6.293, 9.734) &                    0.36 (3.05) &                  -0.12 (-0.99) \\
              Mkt-RF &     2,505.58 &              0.263 &    0.7 & 4.477 &   -0.654 &    2.202 & (-7.212, 7.191) &                  -0.68 (-3.24) &                    0.75 (3.92) \\
                 SMB &       264.59 &             -0.160 &  0.223 & 2.914 &    0.207 &     3.83 & (-4.111, 4.740) &                    0.34 (0.62) &                  -0.31 (-0.56) \\
                 HML &       220.58 &             -0.199 &   0.19 & 2.948 &    0.079 &    2.463 & (-4.082, 5.213) &                    0.58 (3.29) &                  -0.40 (-2.05) \\
                 RMW &       467.17 &             -0.070 &  0.312 & 2.279 &     -0.4 &   12.862 & (-2.583, 3.482) &                    0.07 (0.32) &                    0.07 (0.30) \\
                 CMA &       318.83 &             -0.222 &  0.234 & 1.936 &    0.418 &    2.069 & (-2.521, 3.284) &                  -0.58 (-2.87) &                    0.68 (3.80) \\
                 MOM &     1,547.36 &              0.199 &  0.608 & 4.366 &   -1.335 &   10.738 & (-6.555, 6.701) &                  -0.41 (-1.51) &                    0.49 (1.87) \\
                  RF &       687.29 &                    &  0.358 & 0.291 &    0.737 &    0.351 &  (0.000, 0.870) &                  0.98 (128.39) &                  -0.13 (-2.54) \\
\bottomrule
\end{tabular}

			\caption{Cumulative returns, Sharpe ratios and moments}
	\end{subfigure}
	\end{minipage}
	\\
	\begin{minipage}[h]{0.40\textwidth}
			\centering
			\includegraphics[scale = 0.30]{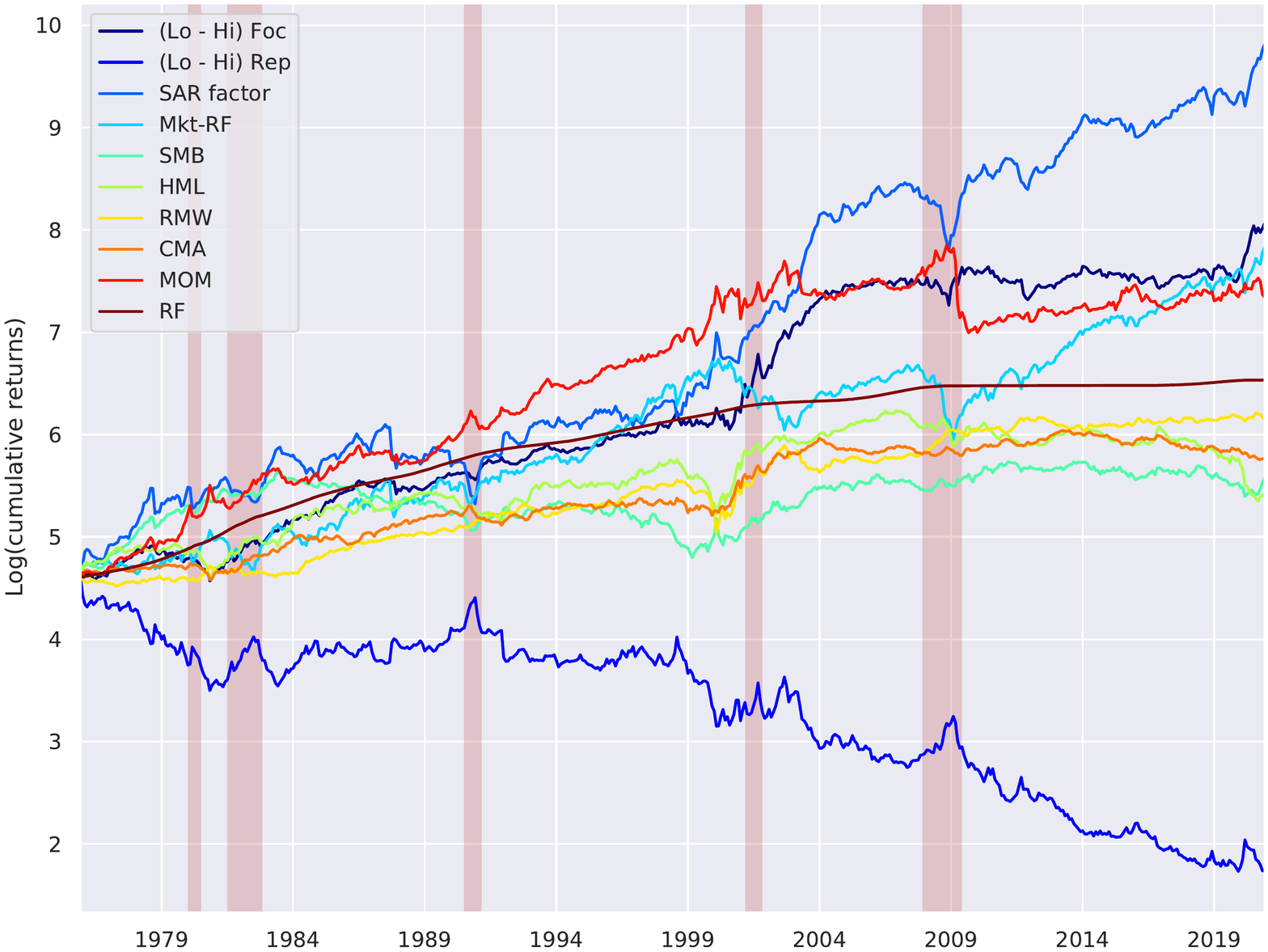}
			\subcaption{Plot of log cumulative returns}
	\end{minipage}
	\hfill
	\begin{minipage}[h]{0.52\textwidth}
		\centering
		\begin{tabular}{lccccccccc}
\toprule
{} &    SAR &  MktRF &    SMB &    HML &   RMW &    CMA &   MOM &  (Lo - Hi) Foc &  (Lo - Hi) Rep \\
\midrule
SAR           &  1.000 &        &        &        &       &        &       &                &                \\
MktRF         &  0.555 &  1.000 &        &        &       &        &       &                &                \\
SMB           &  0.593 &  0.257 &  1.000 &        &       &        &       &                &                \\
HML           & -0.099 & -0.209 & -0.053 &  1.000 &       &        &       &                &                \\
RMW           & -0.385 & -0.259 & -0.399 &  0.198 & 1.000 &        &       &                &                \\
CMA           & -0.107 & -0.359 & -0.051 &  0.674 & 0.113 &  1.000 &       &                &                \\
MOM           & -0.090 & -0.133 &  0.002 & -0.225 & 0.098 & -0.018 & 1.000 &                &                \\
(Lo - Hi) Foc &  0.315 & -0.361 &  0.025 &  0.142 & 0.107 &  0.285 & 0.181 &          1.000 &                \\
(Lo - Hi) Rep & -0.752 & -0.790 & -0.559 &  0.193 & 0.448 &  0.301 & 0.213 &          0.389 &          1.000 \\
\bottomrule
\end{tabular}

		\subcaption{Correlations}
	\end{minipage}
	\caption{\textbf{Cumulative returns and various descriptive statistics of the SAR factor and other assets}. The column ``Cum.\ return'' of Table (a) shows the cumulative returns from an initial $\$100$ investment on December 31, 1975 to December 31, 2020 (the final amount is not inflation adjusted). The column ``Ann.\ Sharpe ratio'' shows the average monthly excess returns of various strategies divided by its standard deviation and then multiplied by $\sqrt{12}$. The third to sixth columns of (a) show the first four moments of the monthly excess returns time series; the mean and standard deviation columns are expressed in percentage points (e.g.\ 1.0 means 1\%). The seventh column shows the 5-th and 95-th percentile of the monthly returns in percentage points. The column ``ARMA(1, $\cdot$) coef'' shows the estimated coefficient $a$ while the column ``ARMA($\cdot$, 1) coef'' shows the estimated coefficient $b$ of the ARMA(1,1) model $r_t = c + a r_{t - 1} + b \epsilon_{t - 1} + \epsilon_t$, and the parentheses show the associated $t$-statistic. Figure (b) shows the log cumulative returns from said initial investment across various investment strategies. The red shaded regions are NBER recession dates. Table (c) shows the time series correlations of our SAR factor against various other factors. The sampling period is monthly from January 31, 1976 to December 31, 2020.
	}
	\label{tab:CumRtn}
\end{table}

\subsection{Additional robustness checks}  
\label{sec:RobustnessChecks}
We run several robustness checks on our main result and they are summarized in Table~\ref{tab:consolidated_robustness_results}. 

\begin{table}[htp!]
			\centering
			\begin{tabular}{llccccccccccc}
\toprule
                  \phantom{Type of result to show here} & \phantom{Return type} &      Lo &       2 &       3 &       4 &       5 &       6 &       7 &       8 &       9 &      Hi & \textit{Lo - Hi} \\
\midrule
(i) \textbf{EN $\mathsf{R^2}$ with min one risky asset} &        \textbf{Focal} &   1.440 &   1.001 &   1.026 &   0.888 &   0.987 &   0.946 &   0.864 &   0.865 &   0.793 &   0.782 &            0.658 \\
                                                        &                       & (5.051) & (3.444) & (3.407) & (3.128) & (3.514) & (3.425) & (3.104) & (3.143) & (2.817) & (2.958) &          (3.901) \\
                                                        &    \textbf{Foc - Rep} &   1.118 &   0.617 &   0.560 &   0.410 &   0.253 &   0.155 &   0.043 &   0.134 &   0.252 &   0.303 &            0.815 \\
                                                        &                       & (4.077) & (2.382) & (2.468) & (1.933) & (1.109) & (0.674) & (0.201) & (0.817) & (2.058) & (2.994) &          (3.147) \\
                                                        &                       &         &         &         &         &         &         &         &         &         &         &                  \\
      (ii) ------------ \textbf{controlling for MktCap} &        \textbf{Focal} &   1.332 &   1.019 &   0.893 &   0.837 &   0.930 &   0.886 &   0.942 &   0.950 &   0.898 &   0.894 &            0.438 \\
                                                        &                       & (5.330) & (4.027) & (3.486) & (3.227) & (3.423) & (3.123) & (3.228) & (3.148) & (2.959) & (2.858) &          (2.826) \\
                                                        &    \textbf{Foc - Rep} &   0.835 &   0.580 &   0.477 &   0.443 &   0.312 &   0.080 &   0.205 &   0.177 &   0.297 &   0.419 &            0.416 \\
                                                        &                       & (3.589) & (2.960) & (2.488) & (2.194) & (1.420) & (0.366) & (1.064) & (1.080) & (2.657) & (3.744) &          (2.163) \\
                                                        &                       &         &         &         &         &         &         &         &         &         &         &                  \\
        (iii) ------------ \textbf{controlling for B/M} &        \textbf{Focal} &   1.380 &   0.988 &   0.998 &   0.952 &   0.981 &   0.912 &   0.916 &   0.931 &   0.912 &   0.888 &            0.492 \\
                                                        &                       & (4.868) & (3.388) & (3.471) & (3.426) & (3.592) & (3.288) & (3.329) & (3.411) & (3.404) & (3.418) &          (3.350) \\
                                                        &    \textbf{Foc - Rep} &   1.068 &   0.596 &   0.543 &   0.442 &   0.259 &   0.122 &   0.101 &   0.201 &   0.289 &   0.396 &            0.672 \\
                                                        &                       & (4.110) & (2.484) & (2.467) & (2.155) & (1.188) & (0.518) & (0.494) & (1.139) & (2.218) & (3.831) &          (2.729) \\
                                                        &                       &         &         &         &         &         &         &         &         &         &         &                  \\
     (iv) ------------ \textbf{controlling for IdioVol} &        \textbf{Focal} &   1.354 &   1.003 &   1.013 &   0.929 &   0.950 &   0.933 &   0.958 &   0.869 &   0.859 &   0.759 &            0.595 \\
                                                        &                       & (5.278) & (3.887) & (3.895) & (3.444) & (3.513) & (3.268) & (3.291) & (2.966) & (2.825) & (2.418) &          (3.669) \\
                                                        &    \textbf{Foc - Rep} &   0.916 &   0.543 &   0.606 &   0.529 &   0.215 &   0.159 &   0.168 &   0.173 &   0.252 &   0.315 &            0.601 \\
                                                        &                       & (3.593) & (2.446) & (2.745) & (2.603) & (0.965) & (0.736) & (0.832) & (1.071) & (2.133) & (2.797) &          (2.672) \\
                                                        &                       &         &         &         &         &         &         &         &         &         &         &                  \\
          (v) ------------ \textbf{controlling for Mom} &        \textbf{Focal} &   1.349 &   1.130 &   1.003 &   0.991 &   0.964 &   0.919 &   0.924 &   0.830 &   0.790 &   0.703 &            0.646 \\
                                                        &                       & (4.794) & (3.955) & (3.464) & (3.540) & (3.463) & (3.333) & (3.384) & (3.042) & (2.809) & (2.628) &          (4.617) \\
                                                        &    \textbf{Foc - Rep} &   1.008 &   0.736 &   0.593 &   0.488 &   0.201 &   0.132 &   0.139 &   0.087 &   0.228 &   0.228 &            0.780 \\
                                                        &                       & (3.771) & (3.068) & (2.782) & (2.238) & (0.954) & (0.570) & (0.638) & (0.488) & (1.891) & (2.278) &          (3.216) \\
                                                        &                       &         &         &         &         &         &         &         &         &         &         &                  \\
         (vi) ------------ \textbf{controlling for STR} &        \textbf{Focal} &   1.435 &   1.008 &   0.925 &   0.934 &   0.954 &   0.849 &   0.949 &   0.879 &   0.830 &   0.809 &            0.626 \\
                                                        &                       & (4.999) & (3.611) & (3.116) & (3.319) & (3.429) & (3.072) & (3.471) & (3.183) & (2.947) & (3.004) &          (3.820) \\
                                                        &    \textbf{Foc - Rep} &   1.116 &   0.606 &   0.486 &   0.445 &   0.247 &   0.005 &   0.133 &   0.148 &   0.302 &   0.322 &            0.795 \\
                                                        &                       & (4.087) & (2.490) & (2.168) & (2.146) & (1.121) & (0.023) & (0.628) & (0.920) & (2.555) & (3.209) &          (3.156) \\
                                                        &                       &         &         &         &         &         &         &         &         &         &         &                  \\
       (vii) ------------ \textbf{controlling for VOLD} &        \textbf{Focal} &   1.259 &   1.010 &   0.921 &   0.873 &   0.959 &   0.871 &   0.971 &   0.929 &   0.958 &   0.928 &            0.331 \\
                                                        &                       & (4.905) & (3.807) & (3.362) & (3.296) & (3.520) & (3.097) & (3.367) & (3.228) & (3.491) & (3.443) &          (2.329) \\
                                                        &    \textbf{Foc - Rep} &   0.800 &   0.544 &   0.534 &   0.507 &   0.223 &   0.132 &   0.197 &   0.200 &   0.383 &   0.410 &            0.390 \\
                                                        &                       & (3.623) & (2.671) & (2.745) & (2.675) & (0.968) & (0.643) & (0.939) & (1.182) & (3.028) & (3.899) &          (2.197) \\
                                                        &                       &         &         &         &         &         &         &         &         &         &         &                  \\
                 (viii) \textbf{FF3 OLS $\mathsf{R^2}$} &        \textbf{Focal} &   1.260 &   1.140 &   1.006 &   0.971 &   0.974 &   0.907 &   0.907 &   0.880 &   0.880 &   0.806 &            0.455 \\
                                                        &                       & (3.961) & (3.667) & (3.245) & (3.131) & (3.363) & (3.197) & (3.421) & (3.497) & (3.482) & (3.165) &          (1.944) \\
                                                        &    \textbf{Foc - Rep} &   1.121 &   0.821 &   0.630 &   0.487 &   0.386 &   0.416 &   0.364 &   0.319 &   0.245 &   0.193 &            0.928 \\
                                                        &                       & (4.353) & (3.787) & (2.998) & (2.546) & (2.087) & (2.490) & (2.092) & (1.749) & (1.399) & (1.216) &          (2.882) \\
                                                        &                       &         &         &         &         &         &         &         &         &         &         &                  \\
      (ix) ------------ \textbf{controlling for MktCap} &        \textbf{Focal} &   0.872 &   0.961 &   0.977 &   1.002 &   1.026 &   0.994 &   0.997 &   1.015 &   0.953 &   0.923 &           -0.052 \\
                                                        &                       & (3.678) & (3.801) & (3.754) & (3.635) & (3.755) & (3.654) & (3.530) & (3.413) & (3.074) & (2.879) &         (-0.274) \\
                                                        &    \textbf{Foc - Rep} &   0.606 &   0.511 &   0.542 &   0.457 &   0.550 &   0.519 &   0.481 &   0.494 &   0.414 &   0.371 &            0.235 \\
                                                        &                       & (3.645) & (3.046) & (3.115) & (2.713) & (3.291) & (3.606) & (3.004) & (3.336) & (2.778) & (2.620) &          (1.672) \\
                                                        &                       &         &         &         &         &         &         &         &         &         &         &                  \\
                    (x) \textbf{FF5 OLS $\mathsf{R^2}$} &        \textbf{Focal} &   1.245 &   1.147 &   1.009 &   0.989 &   0.991 &   0.916 &   0.886 &   0.874 &   0.880 &   0.792 &            0.453 \\
                                                        &                       & (4.029) & (3.612) & (3.237) & (3.213) & (3.396) & (3.246) & (3.350) & (3.482) & (3.509) & (3.097) &          (1.998) \\
                                                        &    \textbf{Foc - Rep} &   1.093 &   0.868 &   0.626 &   0.460 &   0.433 &   0.404 &   0.335 &   0.313 &   0.255 &   0.191 &            0.902 \\
                                                        &                       & (4.484) & (3.838) & (3.058) & (2.280) & (2.502) & (2.340) & (1.927) & (1.729) & (1.461) & (1.247) &          (2.960) \\
                                                        &                       &         &         &         &         &         &         &         &         &         &         &                  \\
      (xi) ------------ \textbf{controlling for MktCap} &        \textbf{Focal} &   0.881 &   0.925 &   1.048 &   1.010 &   1.023 &   0.981 &   0.953 &   1.019 &   0.962 &   0.916 &           -0.035 \\
                                                        &                       & (3.707) & (3.732) & (4.086) & (3.702) & (3.699) & (3.529) & (3.363) & (3.408) & (3.171) & (2.819) &         (-0.182) \\
                                                        &    \textbf{Foc - Rep} &   0.584 &   0.554 &   0.548 &   0.502 &   0.514 &   0.510 &   0.432 &   0.448 &   0.451 &   0.400 &            0.185 \\
                                                        &                       & (3.516) & (3.294) & (3.190) & (2.884) & (3.137) & (3.311) & (2.678) & (3.064) & (3.160) & (2.877) &          (1.360) \\
                                                        &                       &         &         &         &         &         &         &         &         &         &         &                  \\
\bottomrule
\end{tabular}

			\caption{\textbf{Robustness checks}. %
				The reported numbers are monthly returns in percentage points (i.e. 1.0 means 1.0\%) and the parentheses show \cite{Newey1987} robust $t$-statistics with six lags. Result (i) repeats the procedure that shows the main result Table~\ref{tab:univariate_eql_wgt} except we restrict each stock to have a replicate that must contain at least one single risky asset, so the subsetted stocks' replicates cannot simply just be the risk-free asset. Results (ii) to (iv) repeat the bivariate dependent sort procedure of Table~\ref{tab:consolidated_bivariate_eql_wgt} but enforces the aforementioned minimum one risky asset requirement on the stocks. Results (viii) and (x) are analogous to the main result of Table~\ref{tab:univariate_eql_wgt} except we sort stocks by deciles by their FF3 and FF5 $\Rsquared$'s, respectively. FF3 and FF5 replicates for each stock are constructed out of projected normalized OLS coefficients according to Section~\ref{sec:ComparingAgainstFF}. Results (ix) and (xi) show the corresponding FF3 and FF5 $\Rsquared$ decile sort results that first sort stocks into their market capitalization into quintiles; compare analogously to Table~\ref{tab:consolidated_bivariate_eql_wgt} for bivariate elastic-net $\Rsquared$ sort results. The sample period is monthly from January 31, 1976 to December 31, 2020. 
			}
			\label{tab:consolidated_robustness_results}
\end{table} 

\subsubsection{SARP is not driven by the equity risk premium}
\label{sec:MinRiskyStocks}
Is SARP just the equity risk premium in disguise? From Table~\ref{tab:consolidated_characteristics_results}(iv) and the replicate construction procedure of Section~\ref{sec:PortfolioConstruction}, it is evident that stocks with high SAR have replicates that are essentially just the risk-free asset. The SARP for high SAR stocks is simply just the equity premium. However, the SARP for low SAR stocks are distinctively different from their equity premia; there are a plethora of risky peers in the replicates of low SAR stocks. Nonetheless, as constructed, there is potential concern the main empirical message of this paper --- stocks with high (low) SAR have high (low) SARP --- is driven by the equity premium of low SAR stocks. To rule out this concern, we follow the projection and replicate construction procedures of Section~\ref{sec:Methodology}, but then drop all stocks whose replicates only consist of the risk-free assets (that is, the replicates now must contain at least one risky peer). Table~\ref{tab:consolidated_robustness_results}(i) shows the result; compare this against the main result Table~\ref{tab:univariate_eql_wgt}. We see high SAR stocks still have a monthly return of 0.658\% ($t$-stat 3.901) higher than that of low SAR stocks. More importantly, we see the SARP of high SAR stocks (which is now distinctly different from the equity risk premium) is still substantially higher at 0.815\% ($t$-stat 3.147) per month than the SARP of low SAR stocks. Hence the cross-sectional difference in SARP is not driven by the equity risk premium of high SAR stocks. In results (ii) to (vii), we repeat the bivariate sort procedure of Table~\ref{tab:consolidated_bivariate_eql_wgt} and sort various characteristics into quintiles. Our main result remains robust after subsetting for stocks to have at least one risky peer in its replicate, and after controlling for these characteristics.

\subsubsection{Weak SARP when using FF3 and FF5 for replicate construction}
\label{sec:ComparingAgainstFF}
Is there any value at all in running a computationally expensive elastic-net projection procedure of Section~\ref{sec:ProjectionProcedure}? Can we get analogous results using simple OLS projections with much fewer regressors? Instead of projecting each stock $i$ onto the past twelve months' daily returns of every other stock using elastic-net, let's simply project each stock $i$'s past twelve months' daily returns onto the FF3 and FF5 factors using OLS.
\footnote{
	Since we did not project onto an intercept using the elastic-net estimator, we will also not project an intercept when using the FF3 and FF5 OLS regressions \eqref{eq:FF3} and \eqref{eq:FF5}. Recall Remark~\ref{rem:NoIntercept}. 
}
For each stock $i$ and at month $t - 1$, we collect the $3 \times 1$ FF3 OLS projection vector $\boldsymbol{\hat{\beta}}_{i, t - 1}^{\mathrm{FF3}}$ and the $5 \times 1$ FF5 OLS projection vector $\boldsymbol{\hat{\beta}}_{i, t - 1}^{\mathrm{FF5}}$. Symmetric to the normalizing procedure of Section~\ref{sec:PortfolioConstruction}, we will normalize the projection vectors by their $L^1$ norm as $\boldsymbol{\tilde{\beta}}_{i, t - 1}^{\mathrm{FF}z} := \boldsymbol{\hat{\beta}}_{i, t - 1}^{\mathrm{FF}z} / \smallnorm{\boldsymbol{\hat{\beta}}_{i, t - 1}^{\textrm{FF}z}}_1$ for $z = 3, 5$. Analogous to \eqref{eq:Ghost}, the returns of the FF$z$ replicate of stock $i$ is 
\begin{equation}
	R_{i, t}^{\textrm{Rep, FF}z} := (1 - \ind^\top \boldsymbol{\tilde{\beta}}_{i, t - 1}^{\textrm{FF}z}) r_{f,t} + \boldsymbol{R}_{\mathrm{FF}z, t}^\top \boldsymbol{\tilde{\beta}}_{i, t - 1}^{\textrm{FF}z},
	\label{eq:GhostFF}
\end{equation}
where $\boldsymbol{R}_{\mathrm{FF}3, t} = [R_{\mathrm{MktRF}, t}, R_{\mathrm{SMB}, t}, R_{\mathrm{HML}, t}]^\top$ and $\boldsymbol{R}_{\mathrm{FF}5, t} = [R_{\mathrm{MktRF}, t}, R_{\mathrm{SMB}, t}, R_{\mathrm{HML}, t}, R_{\mathrm{CMA}, t}, R_{\mathrm{RMW}, t}]^\top$ are the month $t$ returns of the FF3 and FF5 factors. Finally, we now sort all stocks on their FF$z$ OLS $\Rsquared$'s into deciles. 

Table~\ref{tab:consolidated_robustness_results}(viii) shows the FF3 results while (x) shows the FF5 results. At first glance, it seems we can still identify SARP in the cross-section of FF$z$ $\Rsquared$ decile sorted stocks. The FF3 procedure shows that high FF3 SAR stocks have a SARP of 1.121\% ($t$-stat 4.353) per month while low FF3 SAR stocks have a 0.193\% ($t$-stat 1.216), with a cross-sectional difference of 0.928\% ($t$-stat 2.882). We can see a similar result in the FF5 projections in (iv). However upon closer inspection, the FF$z$ procedures are not robust to size effects, while it is robust for the elastic-net counterpart of Section~\ref{sec:ProjectionProcedure}. In Table~\ref{tab:consolidated_robustness_results}(ix), we repeat the bivariate sort procedure to that of Table~\ref{tab:consolidated_bivariate_eql_wgt} but with FF3 $\Rsquared$ decile bins. After controlling for stocks' market capitalization quintiles, there is no cross-sectional SARP: stocks with high FF3 SAR have a SARP of 0.606\% ($t$-stat 3.645) while stocks with low FF3 SAR have a SARP of 0.371\% ($t$-stat 2.620), but the cross-sectional difference of 0.235\% ($t$-stat 1.672) is statistically insignificant. Likewise in result (xi), we see no cross-sectional SARP in stocks that are constructed out of the aforementioned FF5 projection and sorting procedure, after controlling for size. Overall, these results show SARP can still be found using conventional asset pricing factor models and the standard OLS procedure. However, the ``quality'' of this SARP is rather low compared to the SARP that is identified and constructed via our elastic-net procedure.

\section{Conclusion} 
\label{sec:Conclusion}
This paper theoretically and empirically show the relationship between the \emph{Statistical Arbitrage Risk} (SAR) and \emph{Statistical Arbitrage Risk Premium} (SARP) of a stock. Theoretically, SARP is the expected return of the residual factor risks of a given stock. Empirically we use the elastic-net, a machine learning method, to project a given stock's returns onto the span of every other stock in the market. The projection $\Rsquared$ is SAR, and the normalized projection coefficient entries serve as investment weights in constructing the replicate portfolio of a given stock. The core message of this paper is: \emph{SARP is increasing in SAR}. 

We see several interesting directions to further study SAR and SARP by machine learning (ML). In this paper, the elastic-net serves two simultaneous roles and steps: (1) \emph{fitting and variable selection} and (2) \emph{portfolio construction}. In Step (1), the elastic-net $\Rsquared$ is used to measure the SAR of a stock $i$, and the non-zero entries of the projection vector are used to identify the risky peers of stock $i$. In Step (2), the normalized values of the non-entries are used as investment weights to construct the replicate of stock $i$. A further study of SAR and SARP would be to disentangle these two roles by using two different ML methods. Step (1) can benefit from using a more sophisticated method $\mathrm{ML}_1$ that take in ``wide regressors'' as inputs but has better sparse variable selection and goodness-of-fit $\Rsquared$ (and hence SAR) properties than the elastic-net. For step (2), we see benefits in designing a $\mathrm{ML}_2$ replicate portfolio construction method that also takes into account the relative importance of these peers from step (1). Once these two improved steps are complete, the estimation of SARP and other analyses can follow conventional empirical asset pricing procedures as outlined in this paper. Tangential to using ML methods for forecasting and factors selection in finance, we feel a novel avenue for ML applications in finance is this sense of portfolio construction. In closing, we strongly believe this paper has merely scratched the surface in the study of SAR and SARP by machine learning.

\newpage
\bibliographystyle{ecta} 
\bibliography{main}

\begin{thebibliography}{32}
\newcommand{\enquote}[1]{``#1''}
\expandafter\ifx\csname natexlab\endcsname\relax\def\natexlab#1{#1}\fi

\bibitem[\protect\citeauthoryear{Amihud}{Amihud}{2002}]{Amihud2002}
\textsc{Amihud, Y.} (2002): \enquote{Illiquidity and stock returns:
  cross-section and time-series effects,} \emph{Journal of Financial Markets},
  5, 31--56.

\bibitem[\protect\citeauthoryear{Ang, Hodrick, Xing, and Zhang}{Ang
  et~al.}{2006}]{2006_JF_Ang}
\textsc{Ang, A., R.~J. Hodrick, Y.~Xing, and X.~Zhang} (2006): \enquote{The
  Cross-Section of Volatility and Expected Returns,} \emph{Journal of Finance},
  61, 259--299.

\bibitem[\protect\citeauthoryear{Avellaneda and Lee}{Avellaneda and
  Lee}{2010}]{avellaneda2010statistical}
\textsc{Avellaneda, M. and J.-H. Lee} (2010): \enquote{Statistical arbitrage in
  the US equities market,} \emph{Quantitative Finance}, 10, 761--782.

\bibitem[\protect\citeauthoryear{Bali, Engle, and Murray}{Bali
  et~al.}{2016}]{bali2016empirical}
\textsc{Bali, T.~G., R.~F. Engle, and S.~Murray} (2016): \emph{Empirical Asset
  Pricing: The Cross Section of Stock Returns}, John Wiley \& Sons.

\bibitem[\protect\citeauthoryear{Black, Jensen, and Scholes}{Black
  et~al.}{1972}]{Black1972}
\textsc{Black, F., M.~C. Jensen, and M.~Scholes} (1972): \enquote{The capital
  asset pricing model: Some empirical tests,} \emph{Studies in the theory of
  capital markets}, 81, 79--121.

\bibitem[\protect\citeauthoryear{Breeden, Gibbons, and Litzenberger}{Breeden
  et~al.}{1989}]{Breeden1989}
\textsc{Breeden, D.~T., M.~R. Gibbons, and R.~H. Litzenberger} (1989):
  \enquote{Empirical tests of the consumption-orientated CAPM,} \emph{Journal
  of Finance}, 44, 231--262.

\bibitem[\protect\citeauthoryear{Chinco, Clark-Joseph, and Ye}{Chinco
  et~al.}{2019}]{2018_JF_Chinco}
\textsc{Chinco, A., A.~D. Clark-Joseph, and M.~Ye} (2019): \enquote{Sparse
  signals in the cross-section of returns,} \emph{The Journal of Finance}, 74,
  449--492.

\bibitem[\protect\citeauthoryear{Fama and French}{Fama and
  French}{1992}]{Fama1992cross}
\textsc{Fama, E.~F. and K.~R. French} (1992): \enquote{The cross-section of
  expected stock returns,} \emph{Journal of Finance}, 47, 427--465.

\bibitem[\protect\citeauthoryear{Fama and French}{Fama and
  French}{1993}]{1993_JFE_Fama}
---\hspace{-.1pt}---\hspace{-.1pt}--- (1993): \enquote{Common risk factors in
  the returns on stocks and bonds,} \emph{Journal of Financial Economics}, 33,
  3 -- 56.

\bibitem[\protect\citeauthoryear{Fama and French}{Fama and
  French}{2015}]{Fama2015}
---\hspace{-.1pt}---\hspace{-.1pt}--- (2015): \enquote{A five-factor asset
  pricing model,} \emph{Journal of Financial Economics}, 116, 1 -- 22.

\bibitem[\protect\citeauthoryear{Fama and MacBeth}{Fama and
  MacBeth}{1973}]{fama1973risk}
\textsc{Fama, E.~F. and J.~D. MacBeth} (1973): \enquote{Risk, return, and
  equilibrium: Empirical tests,} \emph{Journal of political economy}, 81,
  607--636.

\bibitem[\protect\citeauthoryear{Feng, Giglio, and Xiu}{Feng
  et~al.}{2020}]{2017_WP_Feng}
\textsc{Feng, G., S.~Giglio, and D.~Xiu} (2020): \enquote{Taming the factor
  zoo: A test of new factors,} \emph{The Journal of Finance}, 75, 1327--1370.

\bibitem[\protect\citeauthoryear{Freyberger, Neuhierl, and Weber}{Freyberger
  et~al.}{2017}]{2018_WP_Freyberger}
\textsc{Freyberger, J., A.~Neuhierl, and M.~Weber} (2017): \enquote{Dissecting
  Characteristics Nonparametrically,} Working Paper 23227, National Bureau of
  Economic Research.

\bibitem[\protect\citeauthoryear{Gatev, Goetzmann, and Rouwenhorst}{Gatev
  et~al.}{2006}]{Gatev2006}
\textsc{Gatev, E., W.~N. Goetzmann, and K.~G. Rouwenhorst} (2006):
  \enquote{Pairs trading: Performance of a relative-value arbitrage rule,}
  \emph{The Review of Financial Studies}, 19, 797--827.

\bibitem[\protect\citeauthoryear{Gu, Kelly, and Xiu}{Gu et~al.}{2020}]{Gu2018}
\textsc{Gu, S., B.~Kelly, and D.~Xiu} (2020): \enquote{Empirical asset pricing
  via machine learning,} \emph{The Review of Financial Studies}, 33,
  2223--2273.

\bibitem[\protect\citeauthoryear{Hansen and Hodrick}{Hansen and
  Hodrick}{1980}]{Hansen1980}
\textsc{Hansen, L.~P. and R.~J. Hodrick} (1980): \enquote{Forward Exchange
  Rates as Optimal Predictors of Future Spot Rates: An Econometric Analysis,}
  \emph{Journal of Political Economy}, 88, 829--853.

\bibitem[\protect\citeauthoryear{Harvey, Liu, and Zhu}{Harvey
  et~al.}{2016}]{Harvey2016}
\textsc{Harvey, C.~R., Y.~Liu, and H.~Zhu} (2016): \enquote{... and the
  Cross-Section of Expected Returns,} \emph{Review of Financial Studies}, 29,
  5--68.

\bibitem[\protect\citeauthoryear{Hedegaard and Hodrick}{Hedegaard and
  Hodrick}{2016}]{Hedegaard2016}
\textsc{Hedegaard, E. and R.~J. Hodrick} (2016): \enquote{Estimating the
  risk-return trade-off with overlapping data inference,} \emph{Journal of
  Banking and Finance}, 67, 135 -- 145.

\bibitem[\protect\citeauthoryear{Huck}{Huck}{2019}]{huck2019large}
\textsc{Huck, N.} (2019): \enquote{Large data sets and machine learning:
  Applications to statistical arbitrage,} \emph{European Journal of Operational
  Research}, 278, 330--342.

\bibitem[\protect\citeauthoryear{Jegadeesh}{Jegadeesh}{1990}]{Jegadeesh1990}
\textsc{Jegadeesh, N.} (1990): \enquote{Evidence of predictable behavior of
  security returns,} \emph{Journal of Finance}, 45, 881--898.

\bibitem[\protect\citeauthoryear{Krauss}{Krauss}{2017}]{Krauss2017}
\textsc{Krauss, C.} (2017): \enquote{Statistical arbitrage pairs trading
  strategies: Review and outlook,} \emph{Journal of Economic Surveys}, 31,
  513--545.

\bibitem[\protect\citeauthoryear{Lamont}{Lamont}{2001}]{Lamont2001}
\textsc{Lamont, O.~A.} (2001): \enquote{Economic tracking portfolios,}
  \emph{Journal of Econometrics}, 105, 161--184.

\bibitem[\protect\citeauthoryear{Lehmann}{Lehmann}{1990}]{Lehmann1990}
\textsc{Lehmann, B.~N.} (1990): \enquote{Fads, martingales, and market
  efficiency,} \emph{Quarterly Journal of Economics}, 105, 1--28.

\bibitem[\protect\citeauthoryear{Leung and Tam}{Leung and
  Tam}{2021}]{leung2021supplement}
\textsc{Leung, R. C.~W. and Y.-M. Tam} (2021): \enquote{Online Supplementary
  Materials for `Statistical Arbitrage Risk Premium by Machine Learning',}
  Tech. rep.

\bibitem[\protect\citeauthoryear{Lintner}{Lintner}{1965}]{Lintner1965}
\textsc{Lintner, J.} (1965): \enquote{The Valuation of Risk Assets and the
  Selection of Risky Investments in Stock Portfolios and Capital Budgets,}
  \emph{The Review of Economics and Statistics}, 47, 13--37.

\bibitem[\protect\citeauthoryear{Newey and West}{Newey and
  West}{1987}]{Newey1987}
\textsc{Newey, W.~K. and K.~D. West} (1987): \enquote{A Simple, Positive
  Semi-Definite, Heteroskedasticity and Autocorrelation Consistent Covariance
  Matrix,} \emph{Econometrica}, 703--708.

\bibitem[\protect\citeauthoryear{Ross}{Ross}{1976}]{Ross1976}
\textsc{Ross, S.} (1976): \enquote{The Arbitrage Theory of Capital Asset
  Pricing,} \emph{Journal of Economic Theory}, 13, 341--360.

\bibitem[\protect\citeauthoryear{Sharpe}{Sharpe}{1964}]{Sharpe1964}
\textsc{Sharpe, W.~F.} (1964): \enquote{Capital Asset Prices: A Theory of
  Market Equilibrium under Conditions of Risk,} \emph{Journal of Finance}, 19,
  425--442.

\bibitem[\protect\citeauthoryear{Shu, Shi, and Tian}{Shu
  et~al.}{2020}]{shu2020high}
\textsc{Shu, L., F.~Shi, and G.~Tian} (2020): \enquote{High-dimensional index
  tracking based on the adaptive elastic net,} \emph{Quantitative Finance},
  1--18.

\bibitem[\protect\citeauthoryear{Tibshirani}{Tibshirani}{1996}]{Tibshirani1996}
\textsc{Tibshirani, R.} (1996): \enquote{Regression shrinkage and selection via
  the lasso,} \emph{Journal of the Royal Statistical Society. Series B
  (Methodological)}, 267--288.

\bibitem[\protect\citeauthoryear{Wurgler and Zhuravskaya}{Wurgler and
  Zhuravskaya}{2002}]{wurgler2002does}
\textsc{Wurgler, J. and E.~Zhuravskaya} (2002): \enquote{Does arbitrage flatten
  demand curves for stocks?} \emph{The Journal of Business}, 75, 583--608.

\bibitem[\protect\citeauthoryear{Zou and Hastie}{Zou and
  Hastie}{2005}]{Zou2005}
\textsc{Zou, H. and T.~Hastie} (2005): \enquote{Regularization and variable
  selection via the elastic net,} \emph{Journal of the Royal Statistical
  Society: Series B (Statistical Methodology)}, 67, 301--320.

\end{thebibliography}
%
\newpage
\appendix 
\addappheadtotoc

\section{Appendix} 
\label{sec:Appendix} 
\subsection{Discussions on the projection procedure} 
\label{sec:EstimationDiscussions} 
Here we outline the details of our projection procedure.

\subsubsection{Elastic-net estimator}
Let's introduce the optimization problem of the \textit{elastic-net} estimator developed by \cite{Zou2005}. Let $y$ be a $T \times 1$ vector, $\mathbf{X}$ be a $T \times N$ matrix and let $\boldsymbol{\beta}$ be a $N \times 1$ vector. Consider the optimization problem, 
\begin{equation}
	\boldsymbol{\hat{\beta}} = \argmin_{\boldsymbol{\beta} \in \R^N}\; \left\{ %
		\frac{1}{2 T} \norm{ y - \mathbf{X}\boldsymbol{\beta} }_2^2 + %
		\lambda_1 \norm{\boldsymbol{\beta}}_1 + %
		\lambda_2 \norm{\boldsymbol{\beta}}_2^2 %
	\right\}, 
	\quad
	\lambda_1, \lambda_2 \ge 0, 
	\label{eq:ElasticNet}
\end{equation}
where $\norm{\mathbf{x}}_1 := \sum_{j=1}^N \abs{x_j}$ is the $L^1$-norm on $\R^N$, and $\norm{\mathbf{x}}_2 := \sqrt{ \sum_{j=1}^N x_j^2 }$ is the $L^2$-norm on $\R^N$. In our application, $y$ will be a time series vector of $T$ days of returns of a particular stock, and $\mathbf{X}$ will be the concatenation of the time series of $N$ number of other stocks. Note this means there are $N + 1$ total number of stocks. 

The solution $\boldsymbol{\hat{\beta}}$ is called the \textit{elastic-net} estimator. This estimator encompasses the special cases of the \textit{ordinary least squares (OLS)} estimator (when $\lambda_1 = 0, \lambda_2 = 0$), \textit{least absolute shrinkage and selection operator (LASSO)} estimator of \cite{Tibshirani1996} (when $\lambda_1 > 0, \lambda_2 = 0$), and the \textit{ridge} estimator (when $\lambda_1 = 0, \lambda_2 > 0$). The hyperparameters $\lambda_1, \lambda_2$ control the strength of the $L^1$- and $L^2$-norm penalties, respectively. In this paper when we refer to the elastic-net estimator, we always refer to the case when $\lambda_1, \lambda_2$ are both strictly positive. In our actual implementation, we use a 3-fold \textit{cross-validation} procedure to empirically select the hyperparameters $\lambda_1, \lambda_2 > 0$.

\begin{rem}
	To reduce the already lengthy computational time in estimating \eqref{eq:ElasticNet}, we cross-validate for only one hyperparameter rather than two in our empirical studies by making the following simplifying assumption on $\lambda_1, \lambda_2$. We set $\lambda_1 = \lambda \ell$ and $\lambda_2 = \frac{1}{2} \lambda (1 - \ell)$, and set $\ell = 1/2$, and only cross-validate for the single $\lambda > 0$ parameter.  
\end{rem}

\subsubsection{Intercept estimation, stale prices and sparsity} 
\label{sec:InterceptEstimation} 
We deliberately do \textit{not} estimate an intercept in \eqref{eq:ElasticNet}.
\footnote{
The elastic-net estimator that contains an intercept is given by,
\begin{equation*}
	\hat{c}, \boldsymbol{\hat{\beta}}
	\in \argmin_{c \in \R, \boldsymbol{\beta} \in \R^p} %
	\left\{ %
		\norm{ y - c - \mathbf{X}\boldsymbol{\beta} }_2^2 + %
		\lambda_1 \norm{\boldsymbol{\beta}}_1 + %
		\lambda_2 \norm{\boldsymbol{\beta}}_2^2 %
	\right\}, 
	\quad
	\lambda_1, \lambda_2 \ge 0, 
\end{equation*}
where by convention, $L^1$- and $L^2$-penalties are only applied on the $\boldsymbol{\beta}$ coefficients and not on the intercept $c$. 
}
This is to avoid attributing a price with very stale prices with high $\Rsquared$. 

Imagine a given stock $i$ has a vector of 12 months' worth daily returns $y_{i,t-1}$, where all the daily returns are almost all $0$'s except on a handful of days. As an extreme, suppose all other stocks has a return matrix $\mathbf{X}_{i,t-1}$ of \eqref{eq:LHSRHSVariables} with rank $D_{i,t-1}$.
\footnote{
	In the actual empirical implementations, we do not impose nor check for any such condition.
}
Then using the elastic-net estimator without intercept \eqref{eq:ElasticNet}, the squared error term would be high, and thus leading to an overall low $\Rsquared_{i,t-1}$. Observe that in this case $\boldsymbol{\hat{\beta}}_{i,t-1} = \boldsymbol{0}$ is not necessarily an optimal solution because $\mathbf{X}_{i,t-1}$ is of rank $D_{i,t-1}$ while $\boldsymbol{\hat{\beta}}_{i, t - 1}$ is $N_{t-1} \times 1$, and we have that $D_{i,t-1} \ll N_{t-1}$. This means there could exist some sparse $\boldsymbol{\hat{\beta}}_{i,t-1} \neq \boldsymbol{0}$ that achieves a smaller value in \eqref{eq:ElasticNet} than that of the zero vector. This is so when $|| y_{i,t-1} - \mathbf{X}_{i,t-1}\boldsymbol{\hat{\beta}}_{i,t-1} ||_2^2 + \lambda_1 ||\boldsymbol{\hat{\beta}_{i,t-1}}||_1 + \lambda_2 ||\boldsymbol{\hat{\beta}}_{i,t-1}||_2^2 \ll || y_{i,t-1} ||_2^2$, especially when the penalty weights $\lambda_1, \lambda_2$ are small. In contrast, if one were to use the elastic-net estimator \textit{with} an intercept, then setting $\boldsymbol{\hat{\beta}}_{i,t-1} = \boldsymbol{0}$  with intercept $\hat{c} \approx 0$ \textit{is} an optimal solution. As a result, this would lead to the numerator term in the calculation of $\Rsquared$ to approximately equal to zero, and thus resulting in a high $\Rsquared$. We want to \textit{avoid} the latter case in our results. 

In this paper, we want to exclusively reserve ``high $\Rsquared$'' to mean stock $i$ can be well explained by other risky assets, and ``low $\Rsquared$'' to mean stock $i$ cannot be explained by other risky assets.

\subsubsection{Parsimony: Why elastic-net and not other machine learning methods?} 
\label{sec:WhyElasticNet} 
Out of a myriad of machine learning methods, why did we choose the elastic-net estimator to test our empirical implication? In Theorem~\ref{thm:ThyReturnDecomp}, we motivated the need to \emph{linearly} regress a stock return $R_i$ onto the returns $\boldsymbol{R}_{-i}$ of all other stocks. The elastic-net estimator can actually be seen as a linear regression problem with constraints. By Lagrange-duality, \eqref{eq:ElasticNet} is identical to the following constrained least squares problem, 
\begin{equation}
\begin{aligned}
	& \min_{\boldsymbol{\beta} \in \R^N }   &     & \frac{1}{2T} \norm{ y - \mathbf{X} \boldsymbol{\beta} }^2_2 \\ 
	& \textrm{subject to} 	    &     & \norm{\boldsymbol{\beta}}_1 \le \eta_1, \\ 
	&                           &     & \norm{\boldsymbol{\beta}}_2^2 \le \eta_2,  
\end{aligned}
\label{eq:ElasticNetConstrained} 
\end{equation}
for some $\eta_1, \eta_2 \ge 0$ that is dependent on the values of $\lambda_1, \lambda_2 \ge 0$. Most references in the statistical literature (e.g.\ \cite{Zou2005}) prefers the unconstrained Lagrangian-form \eqref{eq:ElasticNet} over the constrained optimization form \eqref{eq:ElasticNetConstrained} for theoretical and computational reasons. Here, we explicitly draw out the ``linear nature'' of the elastic-net estimator via \eqref{eq:ElasticNetConstrained} because Theorem~\ref{thm:ThyReturnDecomp} indeed predicts a linear relationship. 

We can now justify why we choose to implement the elastic-net estimator in this paper out of the myriad of machine learning methods. Without a doubt, there are more advanced machine learning and econometric methods that can increase the in-sample fit and thereby boost the in-sample $\Rsquared$. A recent extensive study by \cite*{Gu2018} shows that many possible machine learning methods can achieve high in-sample and out-of-sample $\Rsquared$'s. However, the issue with these ``black box'' methods is that regardless of their in-sample or even out-of-sample performance, their estimated model parameters often do not have a transparent link to the regressors themselves. In contrast, although the elastic-net estimator is inherently non-linear, its estimated coefficients can be applied in linearly back to the regressors. This parsimonious nature of the elastic-net allows us to explicitly and linearly construct the replicates as in \eqref{eq:Ghost}. 

We emphasize that the elastic-net machine learning method is simply a tool to test the prediction of Theorem~\ref{thm:ThyReturnDecomp} via the construction of the replicates \eqref{eq:Ghost}. We have no intentions to conduct statistical inference on the estimated elastic-net coefficients. Our statistical inference claims are still based upon well understood portfolio sort methods in the empirical asset pricing literature as outlined in Section~\ref{sec:PortfolioSort}.   

The consideration of the replicates is what drives us to prefer the elastic-net over its close cousin, the LASSO. The LASSO enjoys a ``sparsity property'', whereby the number of estimated coefficients tend to be small, even though the set of regressors could be large. Sparsity means there are only a handful of stocks that have to be considered in order to construct the replicate of stock $i$. This is important because if the number of stocks needed to construct the replicate of stock $i$ is large, then whatever statistical results we claim to find may be economically infeasible due to trading costs and other market frictions. Unfortunately, if a group of regressors are highly correlated with each other then LASSO has a tendency to only select one regressor effectively at random. This makes for a poor portfolio construction of the replicates for numerous reasons, and loss of potential diversification is an obvious one. The elastic-net remedies this problem by inheriting the grouping property of the ridge estimator. In all, this means the elastic-net is a good candidate for our consideration of the replicates because: (i) it can fit the data (i.e.,\ from the least squares property of the OLS); (ii) it can linearly apply its estimated coefficients over the regressors; (iii) it has a sparsity property (i.e.,\ from LASSO); and (iv) it has a grouping property (i.e.,\ from ridge).

Would a more general machine learning method be useful for our purposes? The answer is mixed. A general form of a machine learning estimator has the form $y = g(\mathbf{X}; \theta, \lambda) + \varepsilon$, where $y$ is the response variable, $g$ could be a parameterized or non-parametric function, $\mathbf{X}$ is the set of regressors, $\theta$ parameterizes $g$, $\lambda$ is a hyperparameter, and $\varepsilon$ is a nuisance parameter. Generally speaking, the relationship between $\mathbf{X}$, $\theta$ and $\lambda$ could be highly non-linear. The data is typically split into three sets $[y_\textrm{train}, y_\textrm{validate}, y_\textrm{test} ]$ and $[\mathbf{X}_\textrm{train}, \mathbf{X}_\textrm{validate}, \mathbf{X}_\textrm{test} ]$. For a given hyperparameter $\lambda$, the machine learning method uses the \emph{training set} $(y_\textrm{train}, \mathbf{X}_\textrm{train})$ to fit the data to find an optimal parameter $\hat{\theta}(\lambda)$. Using the \emph{validation set} $(y_\textrm{valid}, \mathbf{X}_\textrm{valid})$, the method then finds an optimal $\hat{\lambda}$ that fits the validation data, and a trained model is then given by the parameters $\hat{\theta}(\hat{\lambda})$ and $\hat{\lambda}$. Finally, the forecast accuracy of the model is tested against the \emph{testing set} $(y_\textrm{test}, \mathbf{X}_\textrm{test})$ by comparing against the predicted value $\hat{y} = g(\mathbf{X}_\textrm{train} ; \hat{\theta}(\hat{\lambda}), \hat{\lambda} )$ against its realization $y_\textrm{train}$. 

In our context, if we were to apply such general machine learning method to asset $i$, then we would still use the same response variable and regressors as in \eqref{eq:LHSRHSVariables}. However, the fitted value would be of very little use to us when constructing the replicates \eqref{eq:Ghost} from a finance perspective. There are two issues. Firstly, we are \emph{not} using a forecast value $\hat{y}$ to construct portfolios. We are using the fitted coefficients $\hat{\theta}(\lambda) = \boldsymbol{\hat{\beta}}_{i,t-1}$ (which is then subsequently normalized by its $L^1$ norm) as investment weights for the replicate of stock $i$. In order to use fitted coefficients as investment weights, then it is almost \emph{necessary} that these fitted coefficients have a clear and linear relationship to the one-month ahead returns $\mathbf{X}_{\textrm{train}} = \boldsymbol{R}_{-i,t}$. The OLS, LASSO and elastic-net certainly have this property, as seen in \eqref{eq:Ghost}. However, a general machine learning method does \emph{not} have this linear relationship between the fitted coefficients and its out-of-sample regressors. This non-linearity between fitted coefficients and its out-of-sample regressors explicitly prevent us from directly using these general machine learning methods in constructing a portfolio. 

Nonetheless, as discussed at the conclusion of the main text, there are many research avenues to disentangle the fitting role and replicate portfolio construction role of the elastic-net. By designing a method $\mathrm{ML}_1$ for fitting and another method $\mathrm{ML}_2$ for replicate portfolio construction, there are many other properties of the SAR and SARP to explore. Regardless, and at least to us, seems to be the most parsimonious and is a good baseline.



\end{document}